\documentclass[acmsmall,dvipsnames]{acmart}
\acmJournal{PACMPL}
\setcopyright{none}
\renewcommand\footnotetextcopyrightpermission[1]{}
\usepackage{graphicx}
\usepackage{microtype}
\usepackage{enumitem}
\usepackage{listings}
\usepackage{float}
\usepackage{tikzit}
\usepackage{lineno}
\usepackage{subcaption}
\usepackage{mathtools}
\usepackage{tikz}
\usepackage{scalerel}
\usepackage{footmisc}
\usepackage{rotating}
\usepackage{stackengine,xcolor}
\usepackage{array}%
\usepackage{adjustbox}
\usepackage{wasysym}
\usepackage[outline]{contour}
\usetikzlibrary{decorations.pathreplacing,calc}
\tikzstyle{new style 0}=[fill=white, draw=none, shape=rectangle]
\tikzstyle{new style 1}=[fill=white, draw=black, shape=circle]
\tikzstyle{new style 2}=[fill=white, draw=black, shape=rectangle]

\tikzstyle{new edge style 0}=[<-, draw=black, tikzit fill={rgb,255: red,25; green,255; blue,44}, tikzit draw=black]
\tikzstyle{new edge style 1}=[draw=black, {|->}, fill=white, tikzit draw=black]
\tikzstyle{new edge style 2}=[draw=none, fill=none, -, tikzit fill=none]
\tikzstyle{new edge style 3}=[draw={rgb,255: red,125; green,58; blue,71}, <-]
\tikzstyle{new edge style 4}=[<-, draw={rgb,255: red,35; green,66; blue,77}]
\tikzstyle{new edge style 5}=[<->]

\newcommand{\secref}[1]{\autoref{sec:#1}}
\newcommand{\figref}[1]{\autoref{fig:#1}}
\newcommand{\apdxref}[1]{\autoref{appendix:#1}}
\definecolor{bazaar}{rgb}{0.6, 0.47, 0.48}
\definecolor{string}{HTML}{B06500}

\definecolor{typeconstr}{HTML}{4713A9}
\definecolor{obs}{HTML}{23424D}

\newcommand{\Prob}{\mathbb{P}}

\makeatletter
\newcommand*\bigcdot{\mathpalette\bigcdot@{.6}}
\newcommand*\bigcdot@[2]{\mathbin{\vcenter{\hbox{\scalebox{#2}{$\m@th#1\bullet$}}}}}
\makeatother

\makeatletter
\newcommand*\medcdot{\mathpalette\bigcdot@{.4}}
\newcommand*\medcdot@[2]{\mathbin{\vspace*{\fill}{\hbox{\scalebox{#2}{$\m@th#1\bullet$}}}}}
\makeatother
\newcommand*\obsvar[1]{{\color{string}\code{\##1}}}
\newcommand*\cons{\bigcdot}
\newcommand*\concat{\textbf{.}}
\newcommand*\semi{\boldcode{;}}
\newcommand*\assign{\coloneqq}

\newcommand*\LatType{{\color[HTML]{4713A9}\LatConstr}}
\newcommand*\LatConstr{\code{Popl}}% {\code{Latent}}
\newcommand*\ObsType{{\color[HTML]{4713A9}\ObsConstr}}
\newcommand*\ObsConstr{\code{Reported}}%{\code{Observed}}

\newcommand{\tikzmark}[1]{\tikz[overlay,remember picture] \node (#1) {};}

\newcommand*{\AddNote}[5]{%
    \begin{tikzpicture}[overlay, remember picture]
        \draw [decoration={brace,amplitude=0.22em},decorate,semithick,{#5}]
            ($(#3)!([yshift=1.5ex]#1)!($(#3)-(0,1)$)$) --
            ($(#3)!(#2)!($(#3)-(0,1)$)$)
                node [align=center, text width=2.5cm, pos=0.5, anchor=west] {\footnotesize #4};
    \end{tikzpicture}
}%

\newcommand{\dasheduparrow}{\ThisStyle{\vcenter{\hbox{$%
\stackengine{0.45\LMex}{\stackengine{-.15\LMex}{$\SavedStyle\uparrow$}
  {\textcolor{white}{\rule{1.1\LMex}{0.3\LMex}}}{O}{c}{F}{T}{L}%
 }{\textcolor{white}{\rule{1.1\LMex}{0.3\LMex}}}{O}{c}{F}{T}{L}%
$}}}}

\newcommand{\code}[1]{{\small\textsf{#1}}}
\newcommand{\boldcode}[1]{\textbf{\small\textsf{#1}}}

\newcommand*\emptycirc[1][1ex]{\tikz\draw (0,0) circle (#1);}
\newcommand*\halfcirc[1][1ex]{%
  \begin{tikzpicture}
  \draw[fill] (0,0)-- (90:#1) arc (90:270:#1) -- cycle ;
  \draw (0,0) circle (#1);
  \end{tikzpicture}}
\newcommand*\fullcirc[1][1ex]{\tikz\fill (0,0) circle (#1);}

\definecolor{blue(ryb)}{rgb}{0.01, 0.28, 1.0}
\AtBeginDocument{%
  \providecommand\BibTeX{{%
    \normalfont B\kern-0.5em{\scshape i\kern-0.25em b}\kern-0.8em\TeX}}}

\begin{document}

\title{Modular  Probabilistic Models via Algebraic Effects}
\thispagestyle{empty}

\author{Minh Nguyen}
\email{min.nguyen@bristol.ac.uk}
\orcid{0000-0003-3845-9928}
\affiliation{%
  \institution{University of Bristol}
  \city{Bristol}
  \country{United Kingdom}
}
\author{Roly Perera}
\email{rperera@turing.ac.uk }
\orcid{0000-0001-9249-9862}
\affiliation{%
  \institution{The Alan Turing Institute}
  \city{London}
  \country{United Kingdom}
}
\additionalaffiliation{%
   \institution{University of Bristol}
   \city{Bristol}
%   \country{UK}
}
\author{Meng Wang}
\email{meng.wang@bristol.ac.uk }
\orcid{0000-0001-7780-630X}
\affiliation{%
  \institution{University of Bristol}
  \city{Bristol}
  \country{United Kingdom}
}
\author{Nicolas Wu}
\email{n.wu@imperial.ac.uk}
\orcid{0000-0002-4161-985X}
\affiliation{%
  \institution{Imperial College London}
  \city{London}
  \country{United Kingdom}
}

\makeatletter
\let\@authorsaddresses\@empty
\makeatother

\citestyle{acmauthoryear}
\begin{abstract}
  Probabilistic programming languages (PPLs) allow programmers to construct statistical models and then simulate data or perform inference over them. Many PPLs restrict models to a particular instance of simulation or inference, limiting their reusability.  In other PPLs, models are not readily composable. Using Haskell as the host language, we present an embedded domain specific language based on algebraic effects, where probabilistic models are modular, first-class, and reusable for both simulation and inference. We also demonstrate how simulation and inference can be expressed naturally as composable program transformations using algebraic effect handlers.
\end{abstract}


\maketitle

% \input{sec/elaborated-version/main}

% \vspace{-0.25cm}
\section{Introduction}
\label{sec:intro}

In statistics, a probabilistic model captures a real-world phenomenon as a set of relationships between random variables: the model's parameters, inputs, and outputs. By integrating such notions into general-purpose languages, probabilistic programming languages (PPLs) allow programmers to build and execute probabilistic models. For example, consider a simple linear regression model that assumes a linear relationship between input variables $x$ and output variables $y$; this can be represented using the standard mathematical notation shown on the left below. Using the language presented in this paper, the right-hand side shows how one could express the same model as a functional program in Haskell.

\vspace{-0.2cm}
\begin{minipage}{.5\textwidth}
  {\small
  \begin{align*}
    \mu &\sim \mathsf{Normal}(0, 3)\\[-1.5pt]
    c   &\sim \mathsf{Normal}(0, 2)\\[-1.5pt]
    \sigma &\sim \mathsf{Uniform}(1, 3)\\[-1.5pt]
    y &\sim \mathsf{Normal}(\mu * x + c, \, \sigma)
      \end{align*}
}%
\end{minipage}% This must go next to `\end{minipage}`
\begin{minipage}{.5\textwidth}
  \begin{lstlisting}[]
    linRegr |$x$| = do
      |$\mu$| |$\leftarrow$| normal 0 3 |\color{string}\#$\mu$| |\label{line:example-lin-regr-mu}|
      |$c$| |$\leftarrow$| normal 0 2 |\color{string}\#$c$|
      |$\sigma$| |$\leftarrow$| uniform 1 3 |\color{string}\#$\sigma$| |\label{line:example-lin-regr-std}|
      |$y$| |$\leftarrow$| normal (|$\mu\,$|*|$\,x\,$|+|$\,c$|,|$\,\sigma$|) |\color{string}\#$y$| |\label{line:example-lin-regr-y}|
      return |$y$|
    \end{lstlisting}
\end{minipage}

\noindent
Both representations take an input $x$ and specify the distributions which generate the model parameters $\mu$, $c$, and $\sigma$; the output $y$ is then generated from the normal distribution using mean $\mu \, * \, x \, + \, c$ and standard deviation $\sigma$. In the program representation, each primitive distribution is associated with a corresponding  ``observable variable'', indicated by the {\color{string}\lstinline{#}} syntax; this is an optional argument, and its purpose will become clear shortly.

Given a probabilistic model, the programmer or data scientist will typically want to use it in at least two different ways. \emph{Simulation} involves providing fixed values for the model parameters and inputs, to generate the resulting model outputs. Conversely, \emph{inference} generally entails providing observed values for the model outputs and inputs, in an attempt to learn the model parameters.

% What then distinguishes a probabilistic language from a general-purpose one are two effectful operations: \lstinline{sample}, namely drawing a value from a probability distribution, and \lstinline{observe}, which is to incorporate an observation about external data by conditioning a distribution against it \cite{van2018introduction}. Given these two notions, one should then be able to \textit{simulate} outputs from a model, and \textit{infer} the parameters of a model.

\begin{figure}
  % \vspace{-0.1cm}
  \hspace*{-0.4cm}
  \begin{subfigure}{0.375\textwidth}
    \centering
    \includegraphics[width=1\columnwidth]{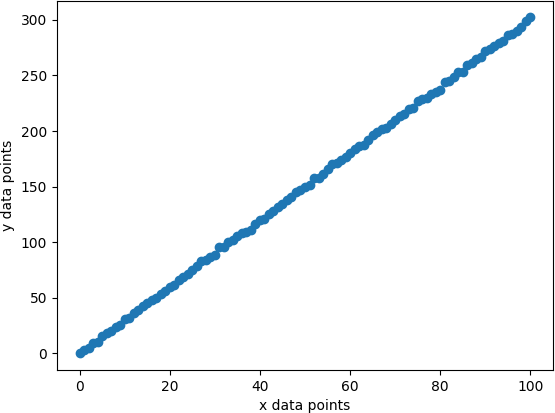}
    \vspace{-0.5cm}
    \caption{Simulation}
    \label{fig:example-lin-regr-sim}
  \end{subfigure}
  \hspace*{0.1cm}
  \begin{subfigure}{0.375\textwidth}
    \center
    \includegraphics[width=1\columnwidth]{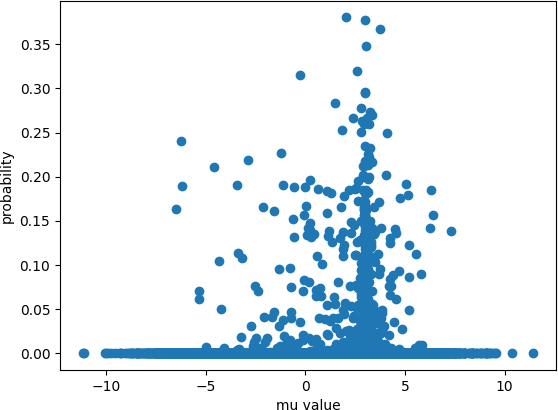}
    \vspace{-0.5cm}
    \caption{Inference (Likelihood Weighting)}
    \label{fig:example-lin-regr-inf}
  \end{subfigure}
  \vspace{-0.3cm}
\caption{Visualising Linear Regression}
\vspace{-0.45cm}
\end{figure}

For example, we might simulate from \lstinline{linRegr} in our language as follows:
\begin{lstlisting}
let |$xs$|  = [ 0 .. 100 ]
    env = (|{\color{string}\#$\mu$} $\coloneqq$ |[3]) |$\cons$| (|{\color{string}\#$c$} $\coloneqq$ |[0]) |$\cons$| (|{\color{string}\#$\sigma$} $\coloneqq$ |[1]) |$\cons$| (|{\color{string}\#$y$} $\coloneqq$ |[|\,|]) |$\cons$| nil
in  map |\,|(simulate linRegr env) |$xs$|
\end{lstlisting}
First we declare a list of model inputs $xs$ from 0 to 100. Then we define a ``model environment'' \lstinline{env} which assigns values \lstinline{3}, \lstinline{0}, and \lstinline{1} to observable variables \obsvar{$\mu$}, \obsvar{$c$}, and \obsvar{$\sigma$}. This expresses our intention to \textit{observe} parameters $\mu$, $c$, and $\sigma$ --- that is, to provide external data $3$ as the value of random variable $\mu$ whilst conditioning on the likelihood that $\mu = 3$, and similarly for $c$ and $\sigma$. On the other hand no values are specified for \obsvar{$y$} in \lstinline{env}, expressing our intent to \textit{sample} the model output $y$ --- that is, to draw a value from its probability distribution. We then use library function \lstinline{simulate} to simulate a single output from the model for each data point in \code{$xs$} using the specified environment, producing the result visualised in \figref{example-lin-regr-sim}.

Alternatively, we can perform inference on \lstinline{linRegr}, for example using the Likelihood Weighting algorithm \cite{van2018introduction}, as follows:
\begin{lstlisting}[escapechar=^] effect
let ^$xs$^  = [ 0 .. 100 ]
    ^$xys$^ = [(^$x$^, env) | ^$x$^ ^$\leftarrow$^ ^$xs$^, let^\,^env = (^\obsvar{$\mu$} $\coloneqq$ ^[])^$\;$^^$\cons$^^$\;$^(^{\color{string}\#$c$} $\coloneqq$ ^[])^$\;$^^$\cons$^^$\;$^(^{\color{string}\#$\sigma$} $\coloneqq$ ^[])^$\;$^^$\cons$^^$\,\,$^(^{\color{string}\#$y$} $\coloneqq$ ^[3^\,^*^\,$x$^])^$\;$^^$\cons$^^$\;$^nil]
in  map ^\,^(lw 200 linRegr^\!^) ^$xys$^
\end{lstlisting}
Here we define $xys$ to pair each model input $x$ with a model environment \code{env} that assigns the value $3 * x$ to \obsvar{$y$} but nothing to \obsvar{$\mu$}, \obsvar{$c$}, and \obsvar{$\sigma$}. This expresses our intention to \textit{observe} $y$ but \textit{sample} $\mu$, $c$ and $\sigma$. We then use library function \lstinline{lw} to perform 200 iterations of Likelihood Weighting for each pair of model input and environment, producing a trace of weighted parameters $\mu$, $c$, and $\sigma$ whose distributions express the most likely parameter values to give rise to $y$. \figref{example-lin-regr-inf} visualises the likelihoods of samples for $\mu$, where values around $\mu = 3$ clearly accumulate higher probabilities.

We refer to a model that can be used for both simulation and inference --- where random variables can be switched between \lstinline{sample} and \lstinline{observe} modes without altering the model itself --- as a \textit{multimodal model}. While multimodal models have a clear benefit, letting the same model be interacted with for a variety of applications, few existing PPLs support them. Most frameworks, such as MonadBayes~\cite{mbayes} and Anglican~\cite{tolpin2016design}, instead require programmers to express models in terms of explicit \lstinline{sample} and \lstinline{observe} operations, which considerably limits their reusability. If the user wishes to interact with the ``same'' model in a new way, they have little choice but to reimplement it with a different configuration of \lstinline{sample} and \lstinline{observe} operations.

% In these languages, performing simulation and inference on the above linear regression model would instead require two versions of the same model to be defined.

Indeed, the number of possible model interpretations extends far beyond the two general scenarios of simulation and inference, potentially including \emph{any} combination of \lstinline{sample} and \lstinline{observe} operations that can be instantiated for a model's random variables. Depending on available data and uncertainty about the model, it is common to explore the model's output space by partially providing model parameters and randomly sampling the rest \cite{kline2016bayesian}, or to alternate between which observable variables are being conditioned on \cite{moon1996expectation}. Ideally, all of these possible scenarios would be expressible with a single multimodal model definition, avoiding the need to define and separately maintain a different version of the model for each use case.

While some PPLs \textit{do} support multimodal models, it is usually difficult or impossible to reuse existing models when creating new ones. Stan~\cite{stan} and WinBUGS~\cite{bugs} provide a bespoke language construct for models with its own distinctive semantics, but as well as lacking high-level programming features beyond those essential to model specification, model definitions are unable to reuse other model definitions. Languages like Turing~\cite{turingjl} and Gen~\cite{genjl} take a different approach, supporting multimodal models as macros that are compiled into functions; although they provide some support for compositionality, neither supports models as first-class values. These modularity limitations are especially significant for hierarchical modelling, where the goal is to  explicitly define a composite model with independently defined sub-models \cite{gelman2006data}.

%A consequence of both approaches is that , and moreover models are not first-class and so cannot be composed or otherwise be manipulated by host language features. These properties are essential for modular development of hierarchical models, where compound models are constructed from independently defined sub-models \cite{gelman2006data}. \todo{[To do (roly): revise paragraph]}

In this paper, we present \textit{ProbFX}: a deeply embedded PPL in Haskell where probabilistic models are modular, first-class, and multimodal. Our solution uses algebraic effects \cite{plotkin2003algebraic} and handlers \cite{plotkin2013handling}, allowing models to be captured as syntax, and their semantics deferred to a choice of ``model environment''. By embedding into a functional language, models can then naturally exist as (first-class) functions and leverage all the abstractions and features of the host \cite{conval2003compiling, gibbons2014folding}.

Our approach uses two key Haskell type abstractions: polymorphic sums for expressing effect signatures, and extensible records for model environments. Both of these can in fact be subsumed by row polymorphism \cite{leijen2005extensible}, and so any language with support for polymorphic rows, such as PureScript \cite{freeman2017purescript}, OCaml \cite{leroy2020ocaml} or Links \cite{hillerstrom2018shallow}, or with a type system powerful enough to express something similar (such as Haskell), should be capable of capturing ProbFX's main features. The other type-level devices we use are ergonomic choices specific to Haskell, and are inessential to the main goals.  %Using Haskell in particular means models can naturally exist as functions and leverage a rich set of abstractions \cite{conval2003compiling, gibbons2014folding}

We begin by giving the necessary background and language overview in \secref{background}. Our contributions are then as follows: %\todo{to do -- verify that contributions are in a good state}:
% \vspace{-0.1cm}
\begin{itemize}[leftmargin=2mm]
  \item We demonstrate the features of our language via a realistic case study with real-world applications: the spread of disease during an epidemic (\secref{case-study}).
  \item % In \secref{ssec:freer} to \secref{ssec:distribution-types}, we present a model-based, general-purpose functional PPL in which models are purely syntactic descriptions of a generative process.
  We present an embedding technique that is novel in using algebraic effects to represent probabilistic models (\secref{embedding}), demonstrated with Haskell as the host language. To the best of our knowledge, ours is the first PPL to support models that are both multimodal and first-class.
  \item We provide a modular, type-safe mechanism for associating observed data to the random variables of a model, determining whether probabilistic operations should be interpreted as \lstinline{sample} or \lstinline{observe}  (\secref{core-handlers}). The same mechanism is used to trace samples for plotting or debugging. %Our solution is novel and uses a combination of type-level tricks, but introduces no type-level innovations of its own.
  \item We present a new approach to the compositional implementation of simulation and inference, using effect handlers to perform modular program transformations on models (\secref{simulation-inference}). We illustrate the approach using Likelihood Weighting \cite{van2018introduction} and Metropolis Hastings \cite{wingate2011lightweight}.
  \item We evaluate our language empirically, considering performance against two state-of-the-art PPLs, and language features supported across a range of modern PPLs (\secref{evaluation}).
\end{itemize}
% \vspace{-0.1cm}
\noindent
We discuss related work in more detail in \secref{conclusion}. However, this is not the first time PPLs have been explored with Haskell as a host language. \citet{ErwigK06} implement a probability monad for representing distributions in functional languages; \citet{narayanan2016probabilistic} use a tagless-final embedding \cite{kiselyov2012typed} to encode inference algorithms as type class instances; \citet{mbayes} use monad transformers \cite{liang1995monad} to demonstrate inference as effect composition. Our approach builds on the techniques offered by algebraic effects and extensible data.

The high-level notion of using interpreters to execute the effects of probabilistic models is an established technique in PPLs, and is similar in spirit to algebraic effects. Many PPLs accomplish this through context managers, coroutines, and continuation-passing style transformations \cite{dippl, tolpin2016design, pyro}. However, these approaches fail to delineate between syntax and semantics, preventing models from being interpreted in a fully multimodal fashion. Moreover, the effect-interpreting mechanisms typically operate in weakly-typed, imperative settings, where effects are not associated with types and can occur unrestrictedly in a program. Algebraic effects have the potential to bring a type-safe, compositional discipline to probabilistic programming. There is, however, little existing work in this area. \citet{goldstein:msc} explores the modularity of probabilistic modelling and inference in Koka Bayes: a prototype PPL in the Koka language  \cite{Leijen_2014}, taking advantage of its effect system. The topic has otherwise primarily remained a point of discussion \cite{scibior2015effects, moore2018effect}. We present a novel design and implementation at the intersection of PPLs and algebraic effects.

In addition to the examples in this paper, our embedding has been tested with a range of models implemented in other PPLs, as well as well-known models such as those designed by \citet{gelman2006data}; the full source code is freely available online.\footnote{\url{https://github.com/min-nguyen/prob-fx}}

\section{Background and Language Overview}
\label{sec:background}

% In this section, we give a high-level correspondence between probability theory and PPLs, and investigate the background relationship between how probabilistic models are conventionally expressed mathematically and programmatically; during this, we provide more insight into the desire for first-class, multimodal models and outline what it looks like in our language to achieve this.
A probabilistic model, expressed using the $\sim$ notation introduced in \secref{intro}, describes how a set of random variables are distributed relative to some fixed input. If the model does not condition against any external data, the distribution it describes is the so-called \textit{joint probability distribution}, giving the probabilities of all possible values that its random variables can assume.
For example, the linear regression model in \secref{intro} describes the distribution $\Prob(y, \mu, c, \sigma ; x)$ -- namely, the joint distribution over random variables $y, \mu, c,$ and $\sigma$ given fixed input $x$ as a non-random parameter.

% A probabilistic model, expressed using the $\sim$ notation introduced in \secref{intro}, describes how a set of random variables are distributed according to some fixed input. For example, in the final line $y \sim \textsf{Normal}(\mu * x + c, \sigma)$ of the linear regression model in \secref{intro}, the random variables $\mu$, $c$ and $\sigma$ have been bound to previous distributions, and $x$ occurs as a free variable representing a fixed input. If the model does not condition against any external data, the distribution it describes is the so-called \textit{joint probability distribution}, giving the probabilities of all possible values that its random variables can assume. For example, linear regression in \secref{intro} describes the distribution $\Prob(y, \mu, c, \sigma ; x)$, that is, the joint distribution over random variables $y, \mu, c,$ and $\sigma$ given fixed input $x$ as a non-random parameter.

%The intuition behind this is that (infinitely) sampling from each random variable results in a histogram that corresponds to their joint distribution.

In real-world applications, we typically have known values for only a subset of these random variables, and are interested in how the other variables are distributed with respect to those known values. Consider providing known data $\hat{y}$ for random variable $y$ in linear regression; we say that we \emph{condition on} $y$ having \emph{observed} $\hat{y}$. Using the well-known chain rule for two random variables:
\vspace{-0.05cm}
\begin{equation*}
  \Prob(X, Y) = \Prob(X \; | \; Y) \cdot \Prob(Y)
\end{equation*}

\vspace{-0.1cm}
\noindent
we can derive the resulting distribution as the product $\Prob(\mu, c, \sigma \, | \, y = \hat{y}; x) \cdot \Prob(y = \hat{y})$. The first component, called the \textit{conditional distribution}, describes the probabilities of values for each random variable $\mu$, $c$, $\sigma$ given $y = \hat{y}$ and some input $x$; the second component, called the \textit{prior distribution}, gives the probability that $y$ has value $\hat{y}$. Providing observed data to a probabilistic model can therefore be seen as specialising its joint distribution to some product of a conditional that is favourable to modelling and an associated prior.

%\vspace{-0.1cm}
\vspace{-0.1cm}
\subsection{Multimodal Models}
The chain rule is a powerful tool that allows us to describe a jointly occurring set of events in terms of the variables we will provide data for, and then compute other variables of interest with respect to this; and there are of course as many ways to decompose a joint distribution as there are combinations of variables that can be conditioned on. Since statisticians often have a clear understanding of the variables they wish to learn and those they wish to condition against, models are in practice often specialised to specific conditional distributions (through the chain rule) and expressed as low-level algorithms that explicitly perform sampling and conditioning, such as in \citet{polson2013bayesian, ding2019structural}.

Most PPLs, such as WebPPL \cite{dippl} and Anglican \cite{tolpin2016design}, are then designed to support the direct translation of these low-level model specifications from paper to program via the operations \lstinline{sample} and \lstinline{observe}. These languages are useful for creating model instances tailored to specific situations, but the resulting models are not easy to experiment with. Tasks which should be straightforward, such as exploring random variable behaviours by isolating which ones are sampled from \cite{idreos2015overview} or selectively optimising model parameters \cite{yekutieli2012adjusted}, require alternative specialisations to be created by hand.

%The problem is compounded with more involved endeavours, e.g.~posterior predictive checks \cite{kruschke2013posterior}, or manually propagating the results of a model in simulation mode to a model in inference mode \cite{fushiki2010bayesian}.

\vspace{-0.1cm}
\subsubsection{Multimodal Models via Model Environments} With multimodal PPLs, the programmer specifies a single model which can be used to generate multiple specialisations, representing specific conditional distributions. Such languages require a mechanism for specifying observed data to random variables, determining whether they are to be sampled or observed. For example, Turing.jl lets users choose whether to provide observed values as arguments when invoking a model, with omitting an argument defaulting to sampling \cite{turingjl}; in Pyro, users specify mappings between random variables and observed data via \textit{context managers} that later constrain the values of runtime sampling operations \cite{pyro}. However, these solutions are dynamically typed with no guarantee that the named variables exist or are provided values of the correct type.

Our language supports multimodal models through a novel notion of \emph{model environment}, which we explain in the context of a Hidden Markov Model (HMM) \cite{rabiner1986hmm}:
\begin{figure}[H]
  \vspace{-0.1cm}
  \centering
  % \vspace{1cm}
  \resizebox{0.65\textwidth}{!}{
    \centering\captionsetup[subfigure]{justification=centering}
    \begin{tikzpicture}
	\begin{pgfonlayer}{nodelayer}
		\node [style=new style 1] (0) at (-6.5, -0.5) {$x_1$};
		\node [style=new style 1] (1) at (-3.5, -0.5) {$x_2$};
		\node [style=new style 1] (2) at (-0.5, -0.5) {$x_3$};
		\node [style=new style 2] (3) at (-6.5, 2) {$y_1$};
		\node [style=new style 2] (4) at (-3.5, 2) {$y_2$};
		\node [style=new style 2] (5) at (-0.5, 2) {$y_3$};
		\node [style=none] (6) at (6, 1.75) {\small Observations};
		\node [style=none] (7) at (6, -0.5) {\small Latent States};
		\node [style=none] (8) at (-8.75, -0.5) {};
		\node [style=none] (9) at (1.75, -0.5) {};
		\node [style=none] (10) at (-8.75, -0.5) {};
		\node [style=none] (11) at (-9.75, -0.5) {$\ldots$};
		\node [style=none] (12) at (2.75, -0.5) {$\ldots$};
		\node [style=none] (20) at (-6.5, 1) {};
		\node [style=none] (21) at (-6.5, 0.25) {};
	\end{pgfonlayer}
	\begin{pgfonlayer}{edgelayer}
		\draw [dash pattern=on 1.5pt off 1.5pt, style=new edge style 0, in=90, out=-90] (3) to (0);
		\draw [dash pattern=on 1.5pt off 1.5pt, style=new edge style 0] (4) to (1);
		\draw [dash pattern=on 1.5pt off 1.5pt, style=new edge style 0] (5) to (2);
		\draw [style=new edge style 0] (2) to (1);
		\draw [style=new edge style 0, in=360, out=180] (1) to (0);
		\draw [style=new edge style 0] (0) to (8.center);
		\draw [style=new edge style 0] (9.center) to (2);
	\end{pgfonlayer}
\end{tikzpicture}
  }
  \vspace{-0.3cm}
\end{figure}
\noindent
The idea of a HMM is that we have a series of latent states $x_i$ which are related in some way to observations $y_i$. The HMM is then defined by two sub-models: a transition model ($\rightarrow$) that determines how latent states $x_i$ are transitioned between, and an observation model ($\dasheduparrow$) that determines how $x_i$ is projected to an observation $y_i$. The objective is to learn about $x_i$ given $y_i$.

A simple HMM expressed in typical statistical pseudocode is shown in \figref{hmm-pseudo}; we describe its corresponding implemention in our language in \figref{hmm-prob-fx}:

\begin{figure}[H]
\vspace{-0.25cm}
\hspace{-0.25cm}
\begin{subfigure}{0.37\textwidth}
  \begin{lstlisting}

hmm(|$n, x_0$|)
  |$\Delta_x$| ~ Uniform(0, 1)       |\tikzmark{listing-hmm-param-start}|
  |$\Delta_y$| ~ Uniform(0, 1)       |\tikzmark{listing-hmm-param-end}|
  for |$i$| = |$1 \ldots n$|:
    |$\delta x_i$| ~ Bernoulli(|$\Delta_x$|)    |\tikzmark{listing-trans-start}|
    |$x_i$|  = |$x_{i-1} + \delta x_i$|         |\tikzmark{listing-trans-end}|
    |$y_i$|  ~ Binomial(|$x_i$|, |$\Delta_y$|) |\,||\tikzmark{listing-obs-start} \tikzmark{listing-obs-end}|
  return |$x_n$|
    \end{lstlisting}
\end{subfigure}
\hspace{1.625cm}
% {\color{darkgray}\vline}
\hspace{-1.4cm}
\begin{subfigure}{0.6\textwidth}
        \begin{lstlisting}[escapechar=^]
          hmm :: (Observables env '[^\color[HTML]{B06500}{``$y_i$''}^] Int
                , Observables env '[^\color[HTML]{B06500}{``$\Delta_x$''}^, ^\color[HTML]{B06500}{``$\Delta_{y}$''}^] Double)
              =>  Int -> Int -> Model env es Int
          hmm ^$n$^ ^$x_0$^ = do
            ^$\Delta_x$^ ^$\leftarrow$^ uniform 0 1 ^\color{string}\#$\Delta_x$^
            ^$\Delta_y$^ ^$\leftarrow$^ uniform 0 1 ^\color{string}\#$\Delta_y$^
            let loop ^$i$^ ^$x_{i-1}$^ | ^$i$^ < ^$n$^ = do
                                ^$\delta x_i$^ ^$\leftarrow$^ bernoulli^\!$'$^ ^$\Delta_x$^
                                let ^$x_i$^ = ^$x_{i-1}$^ + ^$\delta x_i$^
                                ^$y_i$^  ^$\leftarrow$^ binomial ^$x_i$^ ^$\Delta_y$^ ^\color{string}\#$y_i$^
                                loop (^$i + 1$^) ^$x_i$^
                          ^\,^ | otherwise = return ^$x_{i-1}$^
            loop 0 ^$x_0$^
    \end{lstlisting}
\end{subfigure}\\
% \hspace{1.5cm}
\vspace{-0.15cm}
\begin{subfigure}{0.48\textwidth}
  \caption{Statistical pseudocode}
  \label{fig:hmm-pseudo}
\end{subfigure}
\hspace{-0.15cm}
% {\color{darkgray}\vline}
% \hspace{0.3cm}
\begin{subfigure}{0.5\textwidth}
  \caption{Implementation}
  \label{fig:hmm-prob-fx}
\end{subfigure}
\vspace{-0.125cm}
\caption{Hidden Markov Model}
\label{fig:hmm-representation}
\vspace{-0.75cm}
\end{figure}

\AddNote{listing-hmm-param-start}{listing-hmm-param-end}{listing-hmm-param-start}{Model parameters}{black}
\AddNote{listing-trans-start}{listing-trans-end}{listing-trans-start}{Transition model}{black}
\AddNote{listing-obs-start}{listing-obs-end}{listing-obs-start}{\; Observation model}{black}

\noindent
The type of \lstinline{hmm} says it is a function that takes two \lstinline{Int}$\!$s as input and returns a \lstinline{Model env es Int}, where \lstinline{env} is the model environment, \lstinline{es} is the effects which the model can invoke (detailed later in \secref{embedding}), and \lstinline{Int} is the type of values the model generates. The constraint \lstinline{Observables} states that \obsvar{$y_i$}\lstinline{::}\lstinline{Int}, \obsvar{$\Delta_x$}\lstinline{::}\lstinline{Double}, and \obsvar{$\Delta_y$}\lstinline{::}\lstinline{Double}, are \textit{observable variables} in the model environment \lstinline{env} which may be conditioned on later when the model is used.

The function \lstinline{hmm} takes the HMM length $n$ and initial latent state $x_0$ as inputs, and specifies the transition and observation parameters $\Delta_x$ and $\Delta_y$ to be distributed uniformly. It then iterates over the $n$ nodes, applying the transition and observation models at each step. The {\color{black}transition model} computes latent state {$x_i$} from state {$x_{i-1}$} by adding a value {$ \delta x_i$} generated from a Bernoulli distribution {{\small$\mathsf{bernoulli}'$}\,$\Delta_x$}. The {\color{black}observation model} generates observation {$y_i$} from {$x_i$} via the distribution {{\small$\mathsf{binomial}$}\,$\;x_i \;\, \Delta_y$}. The final latent state is returned at the end. (Binding the local name $y_i$ is technically redundant here, but emphasises the connection to the $\sim$ notation used by statisticians.)

The hash syntax, for example in {\color{string}\lstinline{#}\small $y_i$}, constructs the unique inhabitant of the type-level string {\color{string}\small ``$y_i$''}, and is how the programmer associates variables in the \lstinline{Observables} \lstinline{env} constraint with specific primitive distributions. They do this to indicate that they may later be interested in providing observed values for these variables to condition on. When they execute a model, they must provide a concrete environment of type \lstinline{env}, and the presence or absence of observed values in that environment will determine whether the distribution tagged with {\color{string}\lstinline{#}\small $y_i$} is to be interpreted as \lstinline{observe} or \lstinline{sample}. The distribution \lstinline{bernoulli}$\!' \; \Delta_x$ has no observable variable, indicating that it is not possible to condition on {$\delta x_i$}; this makes sense because values of {$\delta x_i$} are latent and so it is unlikely that we would ever want to provide data for them. (Primitive distributions like \code{bernoulli} come in primed variants that are always interpreted as \lstinline{sample}.)

A model can then be interpreted as any of its conditioned forms by specifying an appropriate model environment. In our example, the HMM in its unspecialised form represents the joint distribution over its latent states $x_i$, observations $y_i$, and parameters $\Delta_x, \Delta_y$ (given fixed input $x_0$ as the first latent state):
% \vspace{-0.2cm}
\begin{equation*}
  \Prob(x_{1} \ldots x_n, \, y_{1} \ldots y_n, \, \Delta_x, \, \Delta_y; \, x_0)
\end{equation*}
and we can then simulate the HMM (with length $n = 10$ and initial state $x_0 = 0$) by providing values for \obsvar{$\Delta_x$} and \obsvar{$\Delta_y$} in an environment \lstinline{env}:
\begin{lstlisting}
  let |$x_0$| = 0; |$n$| = 10;
      env = (|\color{string}\#$\Delta_x$| |$\assign$| [0.5]) |$\cons$| (|\color{string}\#$\Delta_y$| |$\assign$| [0.8]) |$\cons$| (|\color{string}\#$y_i$| |$\assign$| []) |$\cons$| nil
  in  simulate (hmm |$n$|) |\,| env |$x_0$|
\end{lstlisting}
This indicates that we want to \lstinline{observe} $0.5$ and $0.8$ for $\Delta_x$ and $\Delta_y$, and \lstinline{sample} for each occurrence of $y_i$ (because we provided no values for \obsvar{$y_i$}); there are multiple occurrences of $y_i$ at runtime, one for each $i \in \{1\ldots n\}$, thanks to the iterative structure of the HMM. By the chain rule, the probability density this expresses is:
% \vspace{-0.1cm}
\begin{equation*}
  \Prob(x_1 \ldots x_{10}, \, y_1 \ldots y_{10} \; | \; \Delta_x = 0.5, \, \Delta_y = 0.8; \, x_0 = 0) \cdot \Prob(\Delta_x = 0.5) \cdot \Prob(\Delta_y = 0.8)
\end{equation*}

\vspace{-0.1cm}
\indent In the case of inference, on the other hand, we provide an observation for each $y_i$ and try to learn {$\Delta_x$} and {$\Delta_x$}. This is why (as the reader may already have noticed) a model environment provides a \textit{list} of values for each observable variable, allowing for the situation where the observable variable has multiple dynamic occurrences. In this case we provide 10 observations, one for each $i \in \{1\ldots n\}$:

\begin{lstlisting}
  let |$x_0$| = 0; |$n$| = 10;
      env = (|\color{string}\#$\Delta_x$| |$\assign$| []) |$\cons$| (|\color{string}\#$\Delta_y$| |$\assign$| []) |$\cons$| (|\color{string}\#$y_i$| |$\assign$| [|0, 1, 1, 3, 4, 5, 5, 5, 6, 5|]) |$\cons$| nil
  in  lw 100 (hmm |$n$|) (|$x_0$|, env)
\end{lstlisting}
\noindent
At runtime, the values associated with {\obsvar{$y_i$}} in the model environment are used to condition against the
occurrences of {\obsvar{$y_i$}} that arise during execution, in the order in which they arise. By the chain rule, the probability density expressed by instantiating the model with this environment is:
% \vspace{-0.1cm}
\begin{equation*}
  \Prob(\Delta_x, \, \Delta_y, \, x_1 \ldots x_{10} \; | \; \, y_1 = 0 \ldots y_{10} = 5; \, x_0 = 0) \cdot \Prob(y_1 = 0) \cdot \ldots \cdot \Prob(y_{10} = 5)
\end{equation*}

\noindent
Although the type system ensures that model environments map observable variables to values of an appropriate type, it does not constrain the number of values that are provided. Should observed values run out for a particular variable, any remaining runtime occurrences of the variable will default to \lstinline{sample}; any surplus of values is ignored. While this flexibility could certainly obscure programming errors, other PPLs (such as Turing.jl \cite{turingjl}) take a similar approach, and we also note that the correctness of inference is unaffected. We consider alternative designs in \secref{future-work}.

\subsection{Modular, First-Class Models}
\figref{hmm-pseudo} used a single procedure, written in statistical pseudocode, to express a Hidden Markov Model. Such notations are understood by most mathematicians and are widely used in statistical journals. Even when the model is complex, a monolithic style of presentation prevails, where models are defined from scratch each time rather than built out of reusable components. The design of PPLs such as Stan \cite{stan}, PyMC3 \cite{pymc3} and Bugs \cite{bugs} reflect these non-modular conventions.

%This approach is ideal for formalising models on paper, and it would be contrary to the objective to introduce richer language constructs such as function definitions/composition and non-trivial control-flow.
Programmers, on the other hand, recognise the importance of modularity to maintainability and reusability: they expect to be able to decompose models into meaningful parts. \figref{hmm-pseudo-modular} shows how the programmer may imagine the same HMM as a composition of parts; our language can then support this treatment of models in \figref{hmm-prob-fx-modular}:

%In \figref{hmm-prob-fx}, the transition and observation models are relatively simple, but real-world models can easily become unwieldy unless they are built modularly -- in these situations, we would

%\subsubsection{Modular decomposition of models}
\begin{figure}[H]
\vspace{-0.3cm}
\hspace{-2.5cm}
\begin{subfigure}{0.35\textwidth}
  \begin{lstlisting}

    transModel(|$\Delta_x, x_{i-1}$|)
      |$\delta x_i$| ~ Bernoulli(|$\Delta_x$|)
      |return| |$x_{i-1} + \delta x_i$|
    \end{lstlisting}
\end{subfigure}
\begin{subfigure}{0.55\textwidth}
  \begin{lstlisting}
  transModel :: Double |$\rightarrow$| Int |$\rightarrow$| Model env es Int
  transModel |$\Delta_x$| |$x_{i-1}$| = do
    |$\delta x_i$| |$\leftarrow$| bernoulli|\!$'$| |$\Delta_x$|
    return |$x_{i-1} + \delta x_i$|
    \end{lstlisting}
\end{subfigure}

\hspace{-2.5cm}
\begin{subfigure}{0.35\textwidth}
  \begin{lstlisting}


    obsModel(|$\Delta_y, x_{i}$|)
      |$y_i$| ~ Binomial(|$x_i$|, |$\Delta_y$|)
      |return| |$y_i$|
    \end{lstlisting}
\end{subfigure}
\begin{subfigure}{0.55\textwidth}
  \begin{lstlisting}
  obsModel :: (Observables env '[|\small\color[HTML]{B06500}{``$y_i$''}|] Int|\!|)
              |$\Rightarrow$| Double |$\rightarrow$| Int |$\rightarrow$| Model env es Int
  obsModel |$\Delta_y$| |$x_{i}$| = do
    |$y_i$| |$\leftarrow$| binomial |$x_i$| |$\Delta_y$| |\color{string}\#$y_i$|
    return |$y_i$|
    \end{lstlisting}
\end{subfigure}

\hspace{-2.5cm}
\begin{subfigure}{0.35\textwidth}
  \begin{lstlisting}


    hmmNode(|$\Delta_x, \Delta_y, x_{i-1}$|)
      |$x_i$| ~ transModel(|$\Delta_x, x_{i-1}$|)
      obsModel(|$\Delta_y, x_{i}$|)
      |return| |$x_i$|
    \end{lstlisting}
\end{subfigure}
\begin{subfigure}{0.55\textwidth}
  \begin{lstlisting}
  hmmNode :: (Observables env '[|\small\color[HTML]{B06500}{``$y_i$''}|] Int|\!|)
            |$\Rightarrow$| Double |$\rightarrow$| Double |$\rightarrow$| Int |$\rightarrow$| Model env es Int
  hmmNode |$\Delta_x$| |$\Delta_y$| |$x_{i-1}$| = do
    |$x_i$| |$\leftarrow$| transModel |$\Delta_x$| |$x_{i-1}$|
    obsModel |$\Delta_y$| |$x_{i}$|
    return |$x_i$|
    \end{lstlisting}
\end{subfigure}

\vspace{-0.5cm}
\hspace{-2.5cm}
\begin{subfigure}{0.35\textwidth}
  \begin{lstlisting}



    hmm(|$n, x_0$|)
      |$\Delta_x$| ~ Uniform(0, 1)
      |$\Delta_y$| ~ Uniform(0, 1)
      for |$i$| = |$1 \ldots n$|;
        |$x_i$| ~ hmmNode(|$\Delta_x, \Delta_y, x_{i-1}$|)
      |return| |$x_n$|
    \end{lstlisting}
\end{subfigure}
\begin{subfigure}{0.55\textwidth}
  \begin{lstlisting}
  hmm :: (|\!|Observables env '[|\color[HTML]{B06500}{``$y_i$''}|]|\!| Int|\!|,|\,|Observables env|\!| '[|\color[HTML]{B06500}{``$\Delta_x$''}|,|\color[HTML]{B06500}{``$\Delta_y$''}|]|\!| Double)
        |$\Rightarrow$| Int |$\rightarrow$| Int |$\rightarrow$| Model env es Int
  hmm |$n$| |$x_0$| = do
    |$\Delta_x$| |$\leftarrow$| uniform 0 1 |\color{string}\#$\Delta_x$|
    |$\Delta_y$| |$\leftarrow$| uniform 0 1 |\color{string}\#$\Delta_y$|
    foldl (|$\texttt{>=>}$|) return (|\!\!|replicate |\!\!||$n$| (hmmNode |$\Delta_x$| |$\Delta_y$|)) |\,$x_0$|
  |$\vphantom{aa}$|
    \end{lstlisting}
\end{subfigure}\\
\hspace{-1.5cm}
\begin{subfigure}{0.35\textwidth}
  \vspace{-0.25cm}
  \caption{Statistical Pseudocode}
  \label{fig:hmm-pseudo-modular}
\end{subfigure}
\begin{subfigure}{0.55\textwidth}
  \vspace{-0.25cm}
  \caption{Program Representation}
  \label{fig:hmm-prob-fx-modular}
\end{subfigure}
\vspace{-0.2cm}
\caption{A Modular Hidden Markov Model}
\vspace{-0.3cm}
\label{fig:hmm-prob-fx-modular-overall}
\end{figure}

\noindent
In \figref{hmm-prob-fx-modular} we define the transition and observation distributions as separate models \lstinline{transModel} and \lstinline{obsModel}. These are composed by \lstinline{hmmNode} to define the behaviour of a single node, which is in turn used by \lstinline{hmm} to create a chain of nodes of length \code{$n$}, using \lstinline{replicate} and a fold of Kleisli composition {\small$\texttt{(>=>)}$} to propagate each node's output to the next one in the chain.
\begin{lstlisting}
  (|\texttt{>=>}|) :: (a |$\rightarrow$| Model env es b) |$\rightarrow$| (b |$\rightarrow$| Model env es c) |$\rightarrow$| (a |$\rightarrow$| Model env es c)
\end{lstlisting}
\noindent
The observable variables of \lstinline{hmm} are now inherited from its sub-models: \lstinline{transModel} has none (and so lacks an \lstinline{Observables} constraint entirely), whereas \lstinline{obsModel} declares \obsvar{$y_i$} as its only observable variable.

As well as improving reusability, compositionality also allows programmers to organise models around the structure of the problem domain. For example the functions in \figref{hmm-prob-fx-modular} correspond in a straightforward way to the abstract components of a HMM: \lstinline{transModel} makes it obvious that latent state \code{$x_i$} depends only on the previous state \code{$x_{i-1}$} (and \code{$\Delta_x$}), and \lstinline{obsModel} that observation \code{$y_i$} depends only on the state \code{$x_i$} that produced it (and \code{$\Delta_y$}).

\section{Modular, Multimodal Models: A Case Study}
\label{sec:case-study}

Before we turn to the details of our embedding approach in \secref{embedding}, we present a case study demonstrating our support for modular, first-class, multimodal models. \secref{the-sir-model} introduces our running example, the SIR (Susceptible-Infected-Recovered) model for the spread of disease \cite{liang2021sir}. \secref{extending-the-sir-model} uses the SIR model to show how our language supports higher-order models which can be easily extended and adapted. \secref{multimodal-sir-model} shows how a multimodal model can be used for both simulation and inference in the same application to facilitate Bayesian bootstrapping.

% There are a myriad of further cases in which it is pragmatic to use models this way -- for example, conditioning against higher-dimensional data with a one-dimensional distribution (using $\!$\lstinline{replicate}$\!$), or instantiating models with different parameter sets (using \lstinline{map}).

\subsection{The SIR Model}
\label{sec:the-sir-model}
The SIR model predicts the spread of disease in a fixed population of size $n$ partitioned into three groups: $s$ for \textit{susceptible to infection}, $i$ for \textit{infected}, or $r$ for \textit{recovered} (where $s + i + r = n$). The model tracks how $s$, $i$, and $r$ vary over discrete time $t$ measured in days. Because testing is both incomplete and unreliable, the true $sir$ values for the population cannot be directly observed; however, we can observe the number of reported infections $\xi$. This problem is thus a good fit for a Hidden Markov Model (\secref{background}), where the $sir$ values play the role of latent states of type \LatType{}, and $\xi$ as the observations of type \ObsType{}:

\begin{lstlisting}
  data |\LatType{}|     |\,|= |\LatConstr| { |$s$|::Int|\!|, |$i$|::Int|\!|, |$r$|::Int }
  type |\ObsType{}| = Int
\end{lstlisting}

\noindent
We now show how our language can be used to implement the SIR model as a modular HMM, starting with the transition and observation models.

\paragraph{SIR transition model.}
The transition model describes how the $sir$ values change over a single day; we model two specific dynamics. First, susceptible individuals $s$ transition to infected $i$ at a rate determined by the values of $s$ and $i$ and the contact rate $\beta$ between the two groups. We use a binomial distribution to model each person in $s$ having a $1 - e^{{-\beta i}/{n}}$ probability of becoming infected, and update $s$ and $i$ accordingly:

\begin{lstlisting}s
  trans|$_{si}$| :: Double |$\rightarrow$| |\LatType{}| |$\rightarrow$| Model env es |\LatType{}|
  trans|$_{si}$| |$\beta$| (|\LatConstr{}| |$s$| |$i$| |$r$|) = do
    let |$n$| = |$s$| + |$i$| + |$r$|
    |$\delta si$| |$\leftarrow$| binomial|$'$| |$s$| (1 |$-$| exp ((|$-$||$\beta$| * |$i$|) / |$n$|))
    return (|\LatConstr{}| (|$s$| |$-$| |$\delta si$|) (|$i$| |$+$| |$\delta si$|) |$r$|)
\end{lstlisting}

Second, infected individuals $i$ transition to the recovered group $r$, where a fixed fraction $\gamma$ of people will recover in a given day. Again we use a binomial to model each person in $i$ as having a $1 - e^{-\gamma}$ probability of recovering, and use this to update $i$ and $r$:

\begin{lstlisting}
  trans|$_{ir}$| :: Double |$\rightarrow$| |\LatType{}| |$\rightarrow$| Model env es |\LatType{}|
  trans|$_{ir}$| |$\gamma$| (|\LatConstr{}| |$s$| |$i$| |$r$|) = do
    |$\delta ir$| |$\leftarrow$| binomial|$'$| |$i$| (1 |$-$| exp (|$-$||$\gamma$|))
    return (|\LatConstr{}| |$s$| (|$i$| |$-$| |$\delta ir$|) (|$r$| |$+$| |$\delta ir$|))
\end{lstlisting}

% Here, we first sample $dIR$, i.e. the amount of infected people that recover, according to the binomial distribution where each person in $I$ has a $1 - e^{-\gamma}$ probability of recovering. This is used to accordingly update $I$ and $R$, and the latent state that results from this change is returned.

The overall transition model \lstinline{trans}$_{sir}$ is simply the sequential composition of \lstinline{trans}$_{si}$ and \lstinline{trans}$_{ir}$. Given \code{$\beta$} and \code{$\gamma$} (aggregated into the type \lstinline{TransParams}), \lstinline{trans}$_{sir}$ computes the changes from $s$ to $i$ and then $i$ to $r$ to yield the updated $sir$ population over a single day:
\begin{lstlisting}
  data TransParams = |TransParams| { |$\beta$| :: Double, |$\gamma$| :: Double } |\vspace{0.15cm}|
  trans|$_{sir}$| :: TransParams |$\rightarrow$| |\LatType{}| |$\rightarrow$| Model env es |\LatType{}|
  trans|$_{sir}$| (|TransParams| |$\beta$| |$\gamma$|) = trans|$_{si}$| |$\gamma$| |$\texttt{>=>}$| trans|$_{ir}$| |$\beta$|
\end{lstlisting}

\paragraph{SIR observation model.}
For the observation model, we assume that the reported infections $\xi$ depends only on the number of infected individuals $i$, of which a fixed fraction $\rho$ will be reported. We use the Poisson distribution to model reports occurring with a mean rate of $\rho * i$:

\begin{lstlisting}
  type ObsParams = Double|\vspace{0.15cm}|
  obs|$_{sir}$| :: Observables env '[|\color{string}{``$\xi$''}|] Int |$\Rightarrow$| ObsParams |$\rightarrow$| |\LatType{}| |$\rightarrow$| Model env es |\ObsType{}|
  obs|$_{sir}$| |$\rho$| (|\LatConstr{}| _ |$i$| _) = do
    |$\xi$| |$\leftarrow$| poisson (|$\rho$| * |$i$|) |\color{string}{\#$\xi$}|
    return |$\xi$|
\end{lstlisting}

\noindent
Since we intend this as the observation model, we declare observable variable \obsvar{$\xi$}\lstinline{::Int} in the \lstinline{Observables} constraint and attach it to the Poisson distribution so we can condition on it later.

\paragraph{HMM for the SIR model}

Now the transition and observation models can be combined into a HMM. We build on the modular design in \figref{hmm-prob-fx-modular-overall}, but go a step further by defining a HMM as a higher-order model that is parameterised by sub-models of type \lstinline{TransModel} and \lstinline{ObsModel}:

\begin{lstlisting}
  type TransModel env es ps lat     = ps |$\rightarrow$| lat |$\rightarrow$| Model env es |\!|lat
  type ObsModel   env es ps lat obs = ps |$\rightarrow$| lat |$\rightarrow$| Model env es obs |\vspace{0.15cm}|
\end{lstlisting}
Here \code{ps} represents the types of the model parameters, and \code{lat} and \code{obs} are the types of latent states and observations. The higher-order HMM is then defined as:
\begin{lstlisting}
  hmm :: |\,|Model env es ps|$_1$|          |$\rightarrow$| Model env es ps|$_2$|
      |$\rightarrow$| TransModel env es ps|$_1$| lat |$\rightarrow$| ObsModel env es ps|$_2$| lat |\!|obs
      |$\rightarrow$\!| Int |\!||$\rightarrow$| lat |$\rightarrow$| Model env es |\!|lat
  hmm |\,|transPrior |\,|obsPrior trans obs |$n$| |$x_0$| = do
    |$\theta$| |$\leftarrow$| transPrior
    |$\phi$| |$\leftarrow$| obsPrior
    hmmNode |$x_{i - 1}$| = do |$x_{i}$| |\,$\leftarrow$| |\,|trans |$\theta$| |$x_{i-1}$|
                       |\,$y_{i}$| |$\leftarrow$| obs |$\phi$| |$x_{i}$|
                       |\,\!|return |$x_{i}$|
    |\hspace{-1mm}|foldl (|\texttt{>=>}|) return (|\!\!|replicate |\!$n$| hmmNode) |\,$x_0$|
\end{lstlisting}
\noindent
The input models \lstinline{transPrior} and \lstinline{obsPrior} are first used to generate model parameters \code{$\theta$} and \code{$\phi$}, and \lstinline{trans} and \lstinline{obs} are arbitrary transition and observation models parameterised by \code{$\theta$} and \code{$\phi$} respectively. The last line creates a HMM of length \code{$n$} with initial state \code{$x_0$}.

% The function \lstinline{hmmGen} takes an input model \lstinline{prior}, representing a prior distribution that generates some specified model parameters $\theta$ (in our case, $\beta, \gamma$, and $\rho$), and an arbitrary transition and observation model \lstinline{transModel} and \lstinline{obsModel} parameterised by $\theta$ (in our case, \lstinline{trans}$_{sir}$ and \lstinline{obs}$_{sir}$). It then defines an auxiliary function \lstinline{hmmNode}, which when provided with an input latent state $x_n$, sequentially applies the transition and observation models to yield an output latent state $x_{n+1}$; for us, this would compute the changes in both $sir$ and $\xi$ values over a single day. Lastly, a HMM chain of length $n$ is instantiated using \!\lstinline{replicate}\!, and the initial latent state $x_0$ is propagated through the chain using \lstinline{foldl} {\small\texttt{(>=>)}}, thereby updating it at each HMM node.

\begin{figure}
    \begin{subfigure}{\textwidth}
      \centering
      \includegraphics[width=0.9\columnwidth]{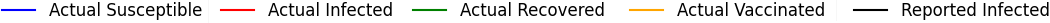}\\
    \end{subfigure}
    \hspace*{-0.4cm}
    \begin{subfigure}{0.325\textwidth}
      \centering
      \includegraphics[width=1.03\columnwidth]{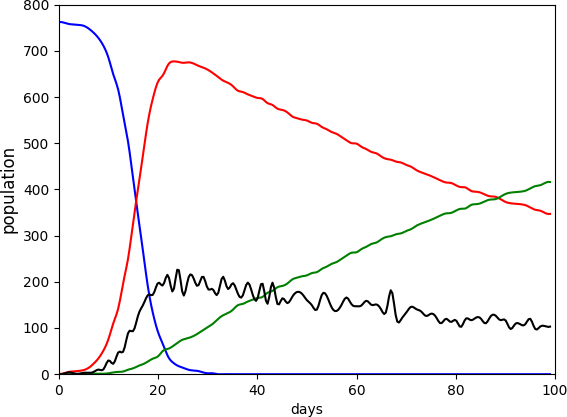}\\
      \vspace{-0.1cm}
      \caption{SIR}
      \label{fig:sir-simulation}
    \end{subfigure}
    \hspace*{0.26cm}
    \begin{subfigure}{0.325\textwidth}
      \centering
      \includegraphics[width=1\columnwidth]{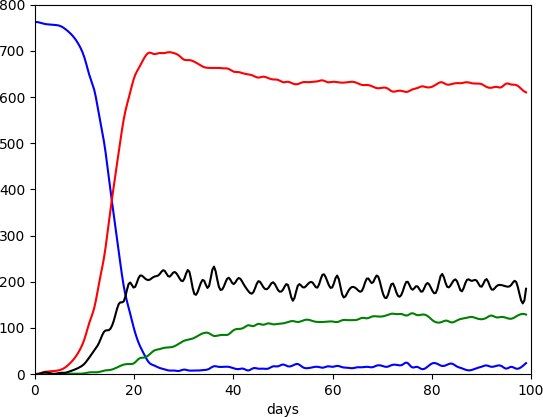}\\
      \vspace{-0.1cm}
      \caption{with resusceptible}
      \label{fig:sir-resusceptible}
    \end{subfigure}
    \hspace*{0.1cm}
    \begin{subfigure}{0.325\textwidth}
      \center
      \includegraphics[width=1\columnwidth]{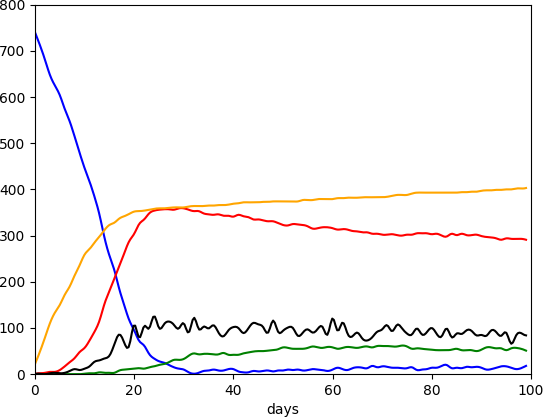}\\
      \vspace{-0.1cm}
      \caption{with resusceptible + vacc}
      \label{fig:sirv-resusceptible}
    \end{subfigure}\\
    \vspace{-0.2cm}
  \caption{SIR Hidden Markov Model Simulation}
  \label{fig:sir}
  \vspace{-0.3cm}
  \end{figure}

Our SIR transition and observation parameters will be provided by models \code{transPrior$_{sir}$} and \code{obsPrior$_{sir}$} below, using primitive distributions \lstinline{gamma} and \lstinline{beta}; their observable variables \obsvar{$\beta$}, \obsvar{$\gamma$}, and \obsvar{$\rho$} will let us condition on those parameters later:

\hspace{-0.5cm}
\begin{minipage}{.5\textwidth}
\begin{lstlisting}
transPrior|\!$_{sir}$| :: (Observables env '[|\color{string}{``$\beta$''}|, |\color{string}{``$\gamma$''}|] |\!|Double)
      |$\Rightarrow$| Model env es TransParams
transPrior|\!$_{sir}$| = do
  |$\beta$| |$\leftarrow$| gamma 2 1 |\color{string}\#$\beta$|
  |$\gamma$| |$\leftarrow$| gamma 1 (1/8) |\color{string}\#$\gamma$|
  return (|TransParams| |$\beta$| |$\gamma$|)
\end{lstlisting}
\end{minipage}
\begin{minipage}{.5\textwidth}
  \begin{lstlisting}[]
      obsPrior|$_{sir}$| :: (Observables env '[|\!\color{string}{``$\rho$''}|] Double)
            |$\Rightarrow$| Model env es ObsParams
      obsPrior|$_{sir}$| = do
        |$\rho$| |$\leftarrow$| beta 2 7 |\color{string}\#$\rho$|
        return |$\rho$|
||
  \end{lstlisting}
\end{minipage}
\noindent
We can now define the complete SIR model. From an initial population \code{$sir$} of susceptible, infected, and recovered individuals, \code{hmm$_{sir}$} models the change in \code{$sir$} over \code{$n$} days given reported infections \code{$\xi$}:
\begin{lstlisting}
  hmm|$_{sir}$| :: (Observables env '[|\color{string}{``$\xi$''}|] Int, Observables env '[|\color{string}{``$\beta$''}|, |\color{string}{``$\gamma$''}|, |\color{string}{``$\rho$''}|] Double)
        |$\Rightarrow$| Int |$\rightarrow$| |\LatType{}| |$\rightarrow$| Model env es |\LatType{}|
  hmm|$_{sir}$| = hmm transPrior|$_{sir}$| obsPrior|$_{sir}$| trans|$_{sir}$| obs|$_{sir}$|
\end{lstlisting}
We can simulate over this model, perhaps to explore some expected model behaviours, by specifying an input model environment \code{sim\_env$_{in}$} of type \lstinline{Env SIRenv} that provides specific values for \obsvar{$\beta$}, \obsvar{$\gamma$}, and \obsvar{$\rho$}, but provides no values for reported infections \obsvar{$\xi$} (ensuring that we always sample for \code{$\xi$}). Applying \lstinline{simulate} to \code{hmm$_{sir}$ 100} and input population \code{$sir_0$} simulates the spread of the disease over 100 days.
\begin{lstlisting}
  type SIRenv = '[ |\color{string}{``$\beta$''}| := Double, |\color{string}{``$\gamma$''}| := Double, |\color{string}{``$\rho$''}| := Double, |\color{string}{``$\xi$''}| := Int ] |\vspace{0.1cm}|
  simulateSIR :: IO (|\LatType{}|, Env SIRenv)
  simulateSIR = do
    let sim_env|$_{in}$|  |\!|= (|{\color{string}\#$\beta$} $\assign$ |[0.7]) |$\cons$| (|{\color{string}\#$\gamma$} $\assign$ |[0.009]) |$\cons$| (|{\color{string}\#$\rho$} $\assign$ |[0.3]) |$\cons$| (|{\color{string}\#$\xi$} $\assign$ |[]) |$\cons$| nil
        |$sir_0$|       = |\LatConstr{}| { |$s$| = 762, |$i$| = 1, |$r$| = 0 }
    simulate (hmm|$_{sir}$| 100) sim_env|$_{in}$| |$sir_0$|
\end{lstlisting}
\noindent
This returns the final population \code{$sir_{100}$} plus an \emph{output} model environment \code{sim\_env$_{out}$} mapping each observable variable to the values sampled for that variable during simulation. From this we can extract the reported infections \code{$\xi$}s:
\begin{lstlisting}
  do (|$sir_{100}$\,|::|\,\LatType{}|, sim_env|$_{out}$||\,|::|\,|Env SIRenv) |$\leftarrow$| simulateSIR
     let |$\xi$|s::|\,|[|\!||\ObsType{}|] = get |{\color{string}\#$\xi$}| sim_env|$_{out}$|
     ...
\end{lstlisting}

\figref{sir-simulation} shows a plot of these $\xi$ values and their corresponding latent (population) states; but note that as it stands, the model provides no external access to the latent states shown in the plot, except the final one \code{$sir_{100}$}. Instrumenting the model to access the intermediate states is discussed in \secref{extending-models-with-effects}.

%To generate this kind of output, it would be ideal if we could access those latent states without polluting the type signature of our HMM abstraction, which concisely corresponds to our statistical definition of a HMM. Fortunately, the setting of algebraic effects means users can orthogonally extend a model with a desired \emph{effect}, such as \lstinline{State}, \lstinline{Reader}, or in this case a \lstinline{Writer} effect for writing the intermediate latent states to a stream; this process is straightforward and shown later in \secref{extending-models-with-effects}.

\subsection{Modular Extensions to the SIR Model}
\label{sec:extending-the-sir-model}

Although the SIR model is simplistic, realistic models may be uneconomical to run or too specific to be useful. When designing models, statisticians aim to strike a balance between complexity and precision, and modular models make it easier to incrementally explore this trade-off. We support this claim by showing how two possible extensions of the SIR model are made easy in our language; while these are by no means the most modular solutions possible, they should suffice to make our point.

%We also show how algebraic effects make it straightforward to extend models with instrumentation and other user-specific effects (\secref{extending-models-with-effects}).

Suppose our disease does not confer long-lasting immunity, so that recovered individuals $r$ transition back to being susceptible $s$ after a period of time \cite{shi2008sis}. We can model this with a new transition behaviour:

\vspace{0.1cm}
\hspace{-2mm}
\begin{minipage}{.45\textwidth}
\begin{lstlisting}
data TransParams = |TransParams| { |$\beta$|::Double, |$\gamma$|::Double, |$\rho$|::Double, |$\eta$|::Double } |\vspace{0.15cm}|
trans|$_{rs}$|:: Double |$\rightarrow$| |\LatType{}| |$\rightarrow$| Model env es |\LatType{}|
trans|$_{rs}$| |$\eta$| (|\LatConstr{}| |$s$| |$i$| |$r$|) = do
  |$\delta rs$| |$\leftarrow$| binomial|$'$| |$r$| (1 - exp (-|$\eta$|))
  return (|\LatConstr{}| (|$s$| + |$\delta rs$|) |$i$| (|$r$| - |$\delta rs$|))
\end{lstlisting}
\end{minipage}% This must go next to `\end{minipage}`
\begin{minipage}{.45\textwidth}
\begin{lstlisting}
|\vspace{0.15cm}|
      trans|$_{sir}$|:: TransModel env es TransParams |\LatType{}|
      trans|$_{sir}$| (|TransParams| |$\beta$| |$\gamma$| |$\eta$|) =
        trans|$_{si}$| |\!$\beta$| |\!\texttt{>=>}\!| trans|$_{ir}$| |\!$\gamma$| |\!\texttt{>=>}\!| trans|$_{rs}$| |\!$\eta$|
||
\end{lstlisting}
\end{minipage}
\vspace{0.1cm}

\noindent
We need only modify \lstinline{TransParams} to include a new parameter \code{$\eta$}; define a new transition sub-model \code{trans$_{rs}$} (parameterised by \code{$\eta$}) that stochastically moves individuals from recovered \code{$r$} to susceptible \code{$s$}; and then adjust \code{trans$_{sir}$} to compose \lstinline{trans}$_{rs}$ with our existing transition behaviours. A simulation of the resulting system is shown in \figref{sir-resusceptible}.

% This could be seen as introducing a new non-parameter variable
% \paragraph{Vaccinated individuals.}
Now consider adding a variant where susceptible individuals $s$ can become vaccinated \code{$v$} \cite{ameen2020efficient}. This involves adding a new sub-population $v$ to the latent state:

\vspace{0.1cm}
\begin{minipage}{.45\textwidth}
\begin{lstlisting}
data |\LatType{}| = |\LatConstr{}| { |$s$|::Int|\!|, |$i$|::Int|\!|, |$r$|::Int|\!|, |$v$|::Int }
data TransParams = |TransParams| { |$\beta$|::Double, |$\gamma$|::Double, |$\rho$|::Double, |$\eta$|::Double, |$\omega$|::Double } |\vspace{0.15cm}|
trans|$_{sv}$|:: Double |$\rightarrow$| |\LatType{}| |$\rightarrow$| Model env es |\LatType{}|
trans|$_{sv}$| |$\omega$| (|\LatConstr{}| |$s$| |$i$| |$r$| |$v$|) = do
  |$\delta sv$| |$\leftarrow$| binomial|$'$| |$s$| (1 - exp (-|$\omega$|))
  return (|\LatConstr{}| (|$s$| - |$\delta sv$|) |$i$| |$r$| (|$v$| + |$\delta sv$|))
\end{lstlisting}
\end{minipage}% This must go next to `\end{minipage}`
\begin{minipage}{.45\textwidth}
\begin{lstlisting}

  |\vspace{0.15cm}|
      trans|$_{sir}$|:: TransModel env es TransParams |\LatType{}|
      trans|$_{sir}$| (|TransParams| |$\beta$| |$\gamma$| |$\eta$| |$\omega$|) =
        trans|$_{si}$| |$\beta$| |\texttt{>=>}| trans|$_{ir}$| |\!$\gamma$| |\texttt{>=>}|
        trans|$_{rs}$| |$\eta$| |\texttt{>=>}| trans|$_{sv}$| |$\omega$|
\end{lstlisting}
\end{minipage}
\vspace{0.1cm}

\noindent
We add field $v$ to \LatType{} representing vaccinated individuals, and add $\omega$ to \lstinline{TransParams}, determining the rate at which $s$ individuals transition to $v$. The new behaviour is expressed by \lstinline{trans}$_{sv}$ and then composed into a new transition model \lstinline{trans}$_{sir}$. A simulation of this is shown in \figref{sirv-resusceptible}.

% the model \code{hmm$_{sir}$} which inherits the \lstinline{Writer [}\LatType{}\lstinline{]} effect from \code{trans$_{sir}$} is composed with the \code{handle$_\textsf{Writer}$} handler, yielding a new SIR model that now also returns a trace of \code{$sir$} values.
% Using this in the context of simulation or inference then follows the same pattern as has been shown in the many previous examples.

\subsection{Exploring Multimodality in the SIR Model}
\label{sec:multimodal-sir-model}

We now show how our support for higher-order, modular models is complemented by the flexibility of multimodal models. Suppose the goal were to infer SIR model parameters $\beta, \gamma, \rho$ given data on reported infections $\xi$. Ideally we would have a real dataset of reported infections to condition on. But what if our dataset were sparse, or if we were interested in quick hypothesis testing? A common option is to use simulated data as observed data, a method called Bayesian bootstrapping \cite{fushiki2010bayesian}. This task is made simple with multimodal models, because we can take the outputs from simulation over a model and plug them into an environment that specifies inference over the same model:

\begin{lstlisting}
  inferSIR :: Env SIRenv |$\rightarrow$| IO (Env SIRenv)
  inferSIR sim_env|$_{out}$| = do
    let |$\xi$s|        = get |\obsvar{$\xi$}| sim_env|$_{out}$|
        mh_env|$_{in}$| = (|{\color{string}\#$\beta$} $\assign$ |[]) |$\cons$| (|{\color{string}\#$\gamma$} $\assign$ |[0.0085]) |$\cons$| (|{\color{string}\#$\rho$} $\assign$ |[]) |$\cons$| (|{\color{string}\#$\xi$} $\assign$| |$\xi$|s) |$\cons$| nil
        |$sir_0$|       |\!|= |\LatConstr{}| { |$s$| = 762, |$i$| = 1, |$r$| = 0 }
    mh 50000 (hmm|$_{sir}$| 100) (|$sir_0$|, mh_env|$_{in}$|)
\end{lstlisting}
Here we take the output model environment \code{sim\_env$_{out}$} produced by \code{simulateSIR}, and use its \code{$\xi$} values to define an input model environment \code{mh\_env$_{in}$} that conditions against \obsvar{$\xi$}. Moreover, suppose we already have some confidence about a particular model parameter, such as the recovery rate $\gamma$; for efficiency, we can avoid inference on $\gamma$ by setting \obsvar{$\gamma$} to an estimate \lstinline{0.0085} and sampling only for the remaining parameters \obsvar{$\beta$} and \obsvar{$\rho$}. We then run Metropolis Hastings \cite{wingate2011lightweight} for \code{50,000} iterations, which returns an output environment (also of type \lstinline{Env SIRenv}) containing the values sampled from all iterations.

\begin{figure}[H]
    \vspace{-0.1cm}
    \hspace{-1cm}
    \begin{subfigure}{0.45\textwidth}
      \center
      \includegraphics[width=0.9\columnwidth]{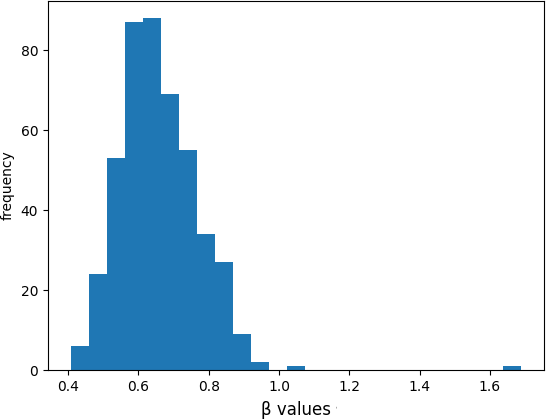}
      \vspace{-0.2cm}
      \caption{SIR $\beta$ posterior distribution}
      \label{fig:sir-beta}
      \vspace{-0.2cm}
    \end{subfigure}
    \begin{subfigure}{0.46\textwidth}
      \center
      \includegraphics[width=0.9\columnwidth]{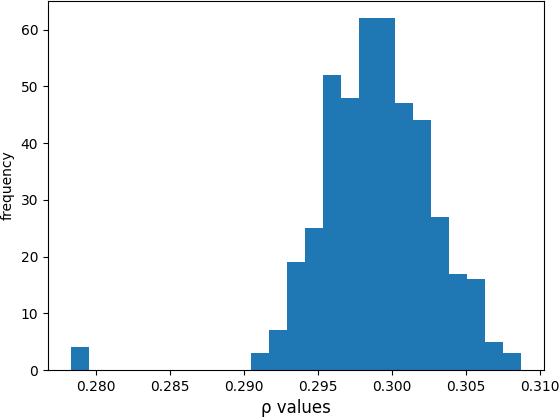}
      \vspace{-0.2cm}
      \caption{SIR $\rho$ posterior distribution}
      \label{fig:sir-rho}
      \vspace{-0.2cm}
    \end{subfigure}
    \caption{SIR Inference (Metropolis Hastings)}
    \label{fig:sir-inf}
    \vspace{-0.1cm}
\end{figure}

The inferred posterior distributions for $\beta$ and $\rho$ can then be obtained simply by extracting them from that environment:
\begin{lstlisting}
  do mh_env|$_{out}$| |$\leftarrow$| inferSIR |$\xi$|s
     let |$\beta$|s = get |{\color{string}\#$\beta$}| mh_env|$_{out}$|
         |$\rho$|s = get |{\color{string}\#$\rho$}| mh_env|$_{out}$|
     |$\ldots$|
\end{lstlisting}
\noindent
These are visualised in \figref{sir-inf}. Values around $\beta = 0.7$ and $\rho = 0.3$ occur more frequently, because these were the parameter values provided in \lstinline{simulateSIR} in \secref{the-sir-model}.

\section{A Syntactic Embedding of Multimodal Models}
\label{sec:embedding}

% To support multimodal models, one of our key design goals is that models should be captured purely syntactically and then assigned semantics upon execution -- this is so we can interpret primitive distributions as either \lstinline{sample} or \lstinline{observe} operations at runtime. However, the behaviour of \lstinline{sample} and \lstinline{observe} further depends on what simulation/inference algorithm the programmer chooses to execute a model with; these algorithms themselves require a variety of effects, such as state for sample tracing and non-determinism for particle simulation, where each effect can orthogonally inject new computational structure to a model. This suggests the desire to compose interpreters of models, and install new effects when necessary.

To support multimodality, a key requirement is that models have a purely syntactic representation which can be assigned a semantics at a later stage. This will allow us to defer the interpretation of primitive distributions as either \lstinline{sample} or \lstinline{observe} operations until we know the variables we want to condition on. An embedding technique that supports this is an \textit{algebraic effect} embedding \cite{plotkin2003algebraic}. In this approach, an effectful program which computes a value of type  \lstinline{a} has the following type:

% a program is typed by some effect signature $es = e_1; \ldots; e_n$, containing the possible computational effects $e$ that it may perform:

\begin{lstlisting}
  prog :: Prog |$es$| a
\end{lstlisting}
The parameter \code{$es$} is called an \emph{effect signature}, in our embedding represented as a type-level list, containing the computational effects that the program may perform. Such programs are syntax trees whose nodes contain operation calls belonging to effect types \code{$e \in es$} and whose leaves contain pure values of type \lstinline{a}.

Programs in this form can be interpreted by \textit{algebraic effect handlers} \cite{plotkin2013handling} which handle specific effects in the program:
\begin{lstlisting}
  handle|$_{\,e}$| :: Prog (|$e: es$|) a |$\rightarrow$| Prog |$es$| b
\end{lstlisting}
Such a handler assigns partial meaning to a program by interpreting all operations \code{$\mathsf{op}$} of type \code{$e$}, discharging \code{$e$} from the effect signature, and transforming the return value of type \lstinline{a} into a value of type \lstinline{b}. Effect handlers are modular building blocks that can be composed into various program interpretations.

When we use this approach to represent a probabilistic computation, nodes of the syntax tree will contain calls to probabilistic operations, and inference and simulation will be implemented as effect handlers that interpret those operation calls. Mechanisms (such as sample tracing or particle simulation) specific to particular simulation or inference algorithms can be supported via effects that inject additional computational structure into the model, such as statefulness or non-determinism.

In this rest of this section (\secref{embedding}), we describe our syntactic representation of multimodal models; providing semantics to this embedding via effect handlers will follow in \secref{core-handlers}.

\subsection{An Infrastructure for Algebraic Effects}
\label{sec:effect-infrastructure}
%Here, a tree (program) structure over operations is then provided by
% \todo{intro}
There are many ways to represent \lstinline{Prog es a}, programs indexed by arbitrary effects; see \citet{hillerstrom2022foundations} for a useful summary. Our embedding uses a simple version of the \textit{freer} monad \cite{freer}:

\begin{lstlisting}
  data Prog es a where
    Val :: |\,|a |$\rightarrow$| Prog es a
    Op  :: forall x. EffectSum es x |$\rightarrow$| (x |$\rightarrow$| Prog es a) |$\rightarrow$| Prog es a
\end{lstlisting}

\noindent
Here, a term of type \lstinline{Prog es a} is a syntax tree whose leaves, \lstinline{Val x}, contain a pure value \lstinline{x} of type \lstinline{a}, and whose nodes, \lstinline{Op op k}, contain an \emph{operation} \lstinline{op} of the abstract datatype \lstinline{EffectSum es x}:

% Here, the effects associated with a model are captured using the following abstract datatype representing an effect sum:
\begin{lstlisting}
  data EffectSum (es :: [Type |$\rightarrow$| Type]) (x :: Type) where |$\ldots$|
\end{lstlisting}

A value of type \lstinline{EffectSum es x} represents a single operation of type \code{e x}, where \lstinline{e} is some effect type constructor that occurs in the type-level list \lstinline{es}, and \lstinline{x} is the type of value the operation produces. The freer monad thus supports multiple effects directly via the constructor \lstinline{Op}, avoiding the need for explicit coproducts (cf.~\citet{swierstra2008data}). Effect sums cannot be constructed directly; their implementation is kept abstract, and the following type class, \lstinline{Member}, is instead provided for working with them:

\begin{lstlisting}[]
  class FindElem e es |$\Rightarrow$| Member e es where
    inj :: e x |$\rightarrow$| EffectSum es x
    prj :: EffectSum es x |$\rightarrow$| Maybe (e x)
\end{lstlisting}

\noindent
The constraint \lstinline{Member e es} asserts that if we can determine \lstinline{e}'s position in \lstinline{es} via the type class \lstinline{FindElem e es} (omitted), then we can safely inject and project an effectful operation, of type \lstinline{e x}, into and out of \lstinline{EffectSum es x}.

The second argument of the \lstinline{Op} constructor is a continuation \lstinline{k} of type \lstinline{x} $\rightarrow$ \lstinline{Prog es a} that takes the result of the operation and constructs the remainder of the syntax tree. This encoding means effect types \lstinline{e} need not be functors, in constrast to approaches based on the free monad. In turn this means the type of distributions (\secref{distribution-effects} below) can be expressed as a GADT, and is in part why we prefer \citeauthor{freer}'s approach to alternatives such as \citet{wu2015fusion} and \citet{kiselyov2013extensible}. Other details of their design specific to performance are omitted for simplicity.

% \cite{freer, polysemy, xie2020effect}.
% effects as \lstinline{m} throughout a computation tree \cite{schrijvers2019monad}, but only in the form of a monad transformer \cite{liang1995monad} which would be unwieldy in comparison to the freer approach.

\subsection{Multimodal Models as Effectful Programs} \label{sec:algebraic-effects-with-ppls}

We now define a probabilistic model, or simply \emph{model}, to be an effectful program of type \lstinline{Prog es a} where \lstinline{es} includes at least two specific effects: \lstinline{Dist} and \lstinline{ObsReader env}. The \textit{distribution} effect \lstinline{Dist} allows the model to make use of primitive distributions such as normal and uniform distributions; the \textit{observable-reader} effect \lstinline{ObsReader env} allows the model to read and update the values of observable variables in a model environment of type \lstinline{env}.

\begin{lstlisting}
  newtype Model env es a =
    |Model| { runModel :: (Member Dist es, Member (ObsReader env) es) |$\Rightarrow$| Prog es a }
\end{lstlisting}
\noindent

\noindent The \lstinline{Member} constraints specify that \lstinline{es} contains \lstinline{Dist} and \lstinline{ObsReader env}. While these two effects suffice for model specification, others may be useful for model execution, and this type allows the model to remain polymorphic in any such additional effects.

\subsubsection{Effects for Distributions}
\label{sec:distribution-effects}

The core computational effect of a probabilistic model is the distribution effect \lstinline{Dist}, allowing models to be formulated in terms of primitive probability distributions. The constructors represent the operations of the effect type, and thus correspond to various primitive distributions. We present a representative selection below:

\begin{lstlisting}
  data Dist a where
    |$\,$|Normal  |\;| :: Double |$\rightarrow$| Double |$\rightarrow$| Maybe Double |$\rightarrow$| Dist Double
    |$\,$|Uniform |\;| :: Double |$\rightarrow$| Double |$\rightarrow$| Maybe Double |$\rightarrow$| Dist Double
    ||Bernoulli :: Double |$\rightarrow$| Maybe Bool |$\rightarrow$| Dist Bool
    ||Discrete  :: [(a, Double)] |$\rightarrow$| Maybe a |$\rightarrow$| Dist a
    |$\ldots$|
\end{lstlisting}

\noindent
Every use of a primitive distribution must, at runtime, be interpretable as either \lstinline{sample} or \lstinline{observe}, depending on the availability of an observed value. Each \lstinline{Dist} operation therefore takes an additional parameter of type \lstinline{Maybe a} (where \lstinline{a} is the base type of the distribution) indicating the presence or absence of a value to condition on. This detail is hidden from the user, but is used internally to determine how operation calls are to be interpreted, as we discuss next.

\subsubsection{Effects for Reading Observable Variables} \label{sec:obsreader-effect} To support conditioning on observable variables via model environments, we require an \textit{observable-reader} effect, \lstinline{ObsReader env}, with an operation that can read and update values of observable variables in \lstinline{env}. This is similar in spirit to the well-known \lstinline{Reader} type.

\begin{lstlisting}
  data ObsReader env a where
    Ask :: Observable env x a |$\Rightarrow$| ObsVar x |$\rightarrow$| ObsReader env (Maybe a)  |\vspace{0.15cm}|
\end{lstlisting}

The single operation \lstinline{Ask} of \lstinline{ObsReader env} has the constraint \lstinline{Observable env x a}. This associates a \emph{list} of values of type \lstinline{a} with observable variable \code{x} in any environment of type \lstinline{env}, supporting conditioning on multiple dynamic instances of the same observable variable. The \lstinline{ObsVar x} argument represents the observable variable name (specified using \obsvar{}). The result is then of type \lstinline{Maybe a} indicating the presence or absence of a value to condition on, suitable for passing directly to one of the \lstinline{Dist} constructors above. We defer the concrete implementations of these types to \secref{model-environments}.

% \subsubsection{Smart Constructors for Multimodal Models}
% \label{sec:smart-constructors-for-distributions}

However, the user is not expected to specify \lstinline{ObsReader} effects directly. Rather, we provide an interface of \textit{smart constructors} \cite{swierstra2008data} which manage the requests to read from observable variables. For example:

\begin{lstlisting}
  |normal |:: (Observable env x Double) |$\Rightarrow$| Double |$\rightarrow$| Double |$\rightarrow$| ObsVar x |$\rightarrow$| Model env es Double
  normal mu sigma x|$'$| = |Model| (do maybe_v |$\leftarrow$| call (Ask x|$'$|)|\semi| |call (Normal mu sigma maybe\_v)|)
\end{lstlisting}
\noindent
Recall that, in addition to the usual normal distribution parameters \lstinline{mu} and \lstinline{sigma}, the constructor \lstinline{Normal} of \lstinline{Dist} also expects an argument of type \lstinline{Maybe Double} representing the presence or absence of an observed value. The smart constructor \lstinline{normal} has a similar signature, but with an observable variable name \code{x$'$} in place of the \lstinline{Maybe Double},
%the rest of this sentence is optional:
thereby associating the distribution with a random variable (rather than particular value) to condition on.
%%%
The role of the smart constructor is to insert an \lstinline{Ask} operation into the program which retrieves the corresponding value \lstinline{maybe_v} of type \lstinline{Maybe Double} from the model environment, which is then used to call the \lstinline{Normal} operation.

The helper function \lstinline{call} simplifies the construction of operation calls, taking care of the injection into \lstinline{EffectSum es x} and supplying the leaf continuation \lstinline{Val}:

\begin{lstlisting}[escapechar=|]
  |call :: |Member e es |$\Rightarrow$| e x |$\rightarrow$| Prog es x
  |call op| = |Op (inj op) Val|
\end{lstlisting}

A primed variant of each smart constructor is also provided, using \lstinline{Nothing} as the observed value, for the common case when a primitive distribution does not need to be conditioned on:

\begin{lstlisting}
  normal|$'$| :: Double |$\rightarrow$| Double |$\rightarrow$| Model env es Double
  normal|$'$| mu sigma = |Model| (|\!|call (Normal mu sigma Nothing))
\end{lstlisting}
\noindent

As an illustration of how the smart constructors work, consider the simple model in \figref{coinflip} which generates a bias \lstinline{p} from a uniform distribution and uses it to parameterise a Bernoulli distribution, determining whether the outcome of a coin flip \lstinline{y} is more likely to be heads (\lstinline{True}) or tails (\lstinline{False}).
\figref{coinflip-program} shows how the user might write the program (omitting the type signature); \figref{coinflip-tree} shows the equivalent program written without smart constructors.

\begin{figure}[H]
    \vspace{-0.2cm}
    % \hspace{-0.5cm}
    \begin{subfigure}{0.45\textwidth}
    \begin{lstlisting}
            coinFlip = do
              p |$\leftarrow$\,|uniform 0 1 |\color{string}{\#p}|
              y |$\leftarrow$|bernoulli p |\color{string}{\#y}|
              return y

              ||
    \end{lstlisting}
    \end{subfigure}
    \hspace{0.2cm}
    % \hspace{0.7cm}
    \begin{subfigure}{0.48\textwidth}
      \begin{lstlisting}
    coinFlip = do
      maybe_p   |$\leftarrow$\,| |\!|call (|Ask| |\obsvar{p}|)
      p         |$\leftarrow$| call (Uniform 0 1 maybe_p)
      maybe_y   |$\leftarrow$\,| |\!|call (|Ask| |\obsvar{y}|)
      y         |$\leftarrow$| call (|\!|Bernoulli p maybe_y)
      return y
      \end{lstlisting}
      % \begin{minipage}{\linewidth}
      % \centering\captionsetup[subfigure]{justification=centering}
      % \tikzfig{figures/tree-1}
      % \end{minipage}
    \end{subfigure}\\
    % \hspace*{-0.5cm}
    \begin{subfigure}{0.4\textwidth}
      \vspace{-0.2cm}
      \caption{User code, with smart constructors}
      \label{fig:coinflip-program}
    \end{subfigure}
    \begin{subfigure}{0.15\textwidth}
    \end{subfigure}
    \hspace{1cm}
    \begin{subfigure}{0.4\textwidth}
      \vspace{-0.2cm}
      \caption{Without smart constructors}
      \label{fig:coinflip-tree}
    \end{subfigure}
    \vspace{-0.2cm}
    \caption{Behaviour of smart constructors}
    \label{fig:coinflip}
    \vspace{-0.3cm}
\end{figure}

%       coinFlip :: ( |\!||\!|Observables env '[|\color{string}{``p''}|] Double
% , |\!||\!|Observables env '[|\color{string}{``y''}|] Bool|\,|)
% |$\Rightarrow$| Model env es Bool

\noindent

\section{Interpreting Multimodal Models}
\label{sec:core-handlers}
We now turn to using effect handlers to assign semantics to models. Interpreting a multimodal model has two stages: specialising the model into the conditional form determined by a model environment, and then executing the resulting specialised model using a particular simulation or inference algorithm.

The first of these stages is the focus of this section. We start with the implementation of model environments (\secref{model-environments}), and then define effect handlers for reading observed values (\secref{handling-obsreader}) and interpreting primitive distributions in response to whether observed values have been provided (\secref{handling-distributions}). Effect handlers for simulation and inference are the topic of \secref{simulation-inference}.

\subsection{Model Environments}
\label{sec:model-environments}
Model environments allow the user to assign values to observable variables which the model can then read from. For the sake of compositionality, models should only need to mention the observable variables they make use of, and be polymorphic in the rest. Moreover, a given observable variable may bind multiple successive values at runtime -- one for each time the variable is evaluated.

These two design constraints suggest a representation of model environments as extensible records, where the fields are observable variable names and the values are \emph{lists} of observed values. We encode these ideas in Haskell as the datatype \lstinline{Env}:

% the correctness and implementation of generic inference algorithms, the execution trace of observed values conditioned against in a probabilistic program must transpire sequentially and in a known order \cite{tolpin2016design}.

\begin{lstlisting}[]
  data Env (env :: [Assign Symbol|\,| Type]) where
    |ENil|   :: Env '[]
    |ECons| :: [a] |$\rightarrow$| Env env |$\rightarrow$| Env ((x := a) : env) |\vspace{0.1cm}|
  data Assign x a = x := a
\end{lstlisting}

\noindent
The type parameter \lstinline{env} represents the type of the model environment as a type-level list of pairs \lstinline{Assign x a}, whose constructor \code{(:=)} associates type-level strings \lstinline{x} of kind \lstinline{Symbol} with value types \lstinline{a} of kind \lstinline{Type}; this tracks the variable names in the environment and their corresponding types. The constructor \lstinline{ENil} is the empty environment, and the constructor \lstinline{ECons} takes a list of values of type \lstinline{a} and an environment of type \lstinline{env} and prepends a new entry for \lstinline{x}, producing an environment of type \lstinline{(x := a) : env}.

The observable variable names in \lstinline{Env}, being type-level strings of kind \lstinline{Symbol}, have no value representation; to use them as record fields at the value-level, we give the singleton datatype \lstinline{ObsVar}:

\begin{lstlisting}[escapechar=|]
  data ObsVar (x :: Symbol) where
    |ObsVar| :: KnownSymbol x |$\Rightarrow$| ObsVar x
\end{lstlisting}

\noindent This acts as a container for \lstinline{Symbol}s, storing them as a phantom parameter \lstinline{x}. String values can be neatly promoted to such containers by deriving an instance of the \lstinline{IsLabel} class, using Haskell's \lstinline{OverloadedLabels} language extension:

\begin{lstlisting}[escapechar=|]
  instance (KnownSymbol x, x |$\sim$| x|$'$|) |$\Rightarrow$| IsLabel x (ObsVar x|$'$|) where
\end{lstlisting}

\noindent
This enables values of type \lstinline{ObsVar} to be created using the {\color{string}\lstinline{#}} syntax, so that for example value {\color{string}\lstinline{#foo}} has type \lstinline{ObsVar}{\;\small\textsf{\color[HTML]{B06500}{``foo''}}}. Model environments are then constructed using the following interface, which provides \lstinline{nil} for the empty environment, and an infix cons-like operator {\small\texttt{($\cons$)}} which makes use of the \lstinline{:=} notation at the value-level:
\begin{lstlisting}
  (|$\cons$|) :: Assign (ObsVar x) [a] |$\rightarrow$| Env env |$\rightarrow$| Env ((x := a) : env)
  nil |\!|:: Env '[]
\end{lstlisting}

Finally, constraining a polymorphic model environment is done via the type class \lstinline{Observable}, which provides type-safe access and updates to observable variables:

\begin{lstlisting}[escapechar=|]
  class (FindElem x env, LookupType x env |$\sim$| a) |$\Rightarrow$| Observable env x a
    get :: ObsVar x |$\rightarrow$| Env env |$\rightarrow$| [a]
    set :: ObsVar x |$\rightarrow$| [a] |$\rightarrow$| Env env |$\rightarrow$| Env env |\vspace{0.15cm}|
  type family LookupType x env where
    LookupType x ((x  |\!|:= a) : env) = a
    LookupType x ((x|$'$| := a) : env) = LookupType x env
\end{lstlisting}
The methods \lstinline{get} and \lstinline{set} use the type class \lstinline{FindElem} (used in \secref{effect-infrastructure}) to find the position of a variable \lstinline{x} in \lstinline{env}, and the type family \lstinline{LookupType} to retrieve its type \lstinline{a}. Many observable variables of the same type can be specified with the type family \lstinline{Observables env xs a} below, which returns a nested tuple of constraints \lstinline{Observable env x a} for each variable \lstinline{x} in \lstinline{xs}:
\begin{lstlisting}
  type family Observables env (xs :: [Symbol])|\,| a :: Constraint where
    Observables env (x : xs) a = (Observable env x a, Observables env xs a)
    Observables env '[] v = ()
\end{lstlisting}

Although other designs are possible, using lists to represent observations for random variables is justified by the fact that, for correctness, general-purpose inference algorithms must compute the same distribution on traces (sequences of sampled or observed values) \cite{tolpin2016design}.

\subsection{Handling Reading of Observable Variables}
\label{sec:handling-obsreader}

The first step in specialising a model to a particular model environment is to handle the \lstinline{Ask} operations of \lstinline{ObsReader}, representing environment read requests.

In the setting of algebraic effects, denoting that an effect \code{e} has been handled is done by \emph{discharging} it from the front of an effect signature \code{e : es}. For this, we introduce a helper function \lstinline{discharge} which pattern-matches an operation of type \lstinline{EffectSum (e : es) a} to determine whether it inhabits the leftmost component \lstinline{e} of the sum:

\begin{lstlisting}
  discharge :: EffectSum (e : es) a |$\rightarrow$| Either (EffectSum|\,| es|\,| a)|\,| (e a)
\end{lstlisting}
If the operation indeed belongs to \lstinline{e}, it is returned as \code{Right op} where \code{op} has type \code{e a}. Otherwise the operation belongs to an effect in \lstinline{es} and it is returned as \code{Left op} where \lstinline{op} has type \lstinline{EffectSum es a}, discharging \lstinline{e} from the effect signature.

We then define a handler which, given a model environment \lstinline{env} and a program with effect signature \lstinline{(ObsReader env : es)}, discharges the \lstinline{ObsReader env} effect by interpreting \lstinline{Ask} operations:

\begin{lstlisting}
  handle|$_\textsf{Read}$| :: Env env |$\rightarrow$| Prog (ObsReader env : es) a |$\rightarrow$| Prog es a
  handle|$_\textsf{Read}$| env (|Op| op k) = case discharge op of
    Right (Ask x) |$\rightarrow$| let |\,|vs       |\,|= |get x env|
                          maybe_v  = |safeHead vs|
                          |env$'$|     = |set x (safeTail vs) env|
                      in  handle|$_\textsf{Read}$| env|$'$| (k maybe_v)
    Left |op$'$|      |$\rightarrow$| |Op| |op$'$| (handle|$_\textsf{Read}$| env |\concat| k)
  handle|$_\textsf{Read}$| env (|Val| x)   = return x
\end{lstlisting}

There are three cases. On matching an operation as \code{Right (Ask x)} containing a request to read from \lstinline{x}, the list of values \lstinline{vs} associated with \lstinline{x} is looked up in \lstinline{env}. If \lstinline{vs} has a head element, contained in \lstinline{maybe_v}, that value becomes the current observation of \lstinline{x} and is removed from \lstinline{env}; this ensures that no observed value is conditioned on more than once during an execution, and that the order in which observations are consumed matches the execution order. Otherwise there is no observation of \lstinline{x} and \lstinline{env} is unchanged. The resulting \lstinline{maybe_v} is provided to the continuation \lstinline{k} to construct the remainder of the program, which is then recursively handled by \lstinline{handle}$_\textsf{Read}$ using the updated environment.

If we match an operation as \code{Left op$'$}, then \code{op$'$} does not belong to \lstinline{ObsReader}. The operation is left intact and the remainder of the program is handled via \code{(handle$_\textsf{Read}$ env \concat{} k)}. Lastly, reaching the non-operation \lstinline{Val x} will simply return value \lstinline{x} as the program's output.

In the handler definitions that follow, we omit the \code{Left op$'$} and \code{Val x} clauses when their implementation follows the pattern above. For an overview of effect handlers we refer the reader to \citet{freer}.
\vspace{-0.1cm}
\subsection{Handling Distributions}
\label{sec:handling-distributions}

The second step of model specialisation is the handler for the \lstinline{Dist} effect, which interprets a primitive distribution call as a sampling or observing operation depending on the presence or absence of an observed value. This requires two new effects, \lstinline{Sample} and \lstinline{Observe}:

\begin{minipage}{.5\textwidth}
\begin{lstlisting}
    data Sample a where
      |Sample|  :: Dist a |$\rightarrow$| Sample a |$\vspace{0.15cm}$|
\end{lstlisting}
\end{minipage}% This must go next to `\end{minipage}`
\begin{minipage}{.5\textwidth}
\begin{lstlisting}
data Observe a where
  |Observe| :: Dist a |$\rightarrow$| a |$\rightarrow$| Observe a  |$\vspace{0.15cm}$|
\end{lstlisting}
\end{minipage}

\noindent
Each has a single operation. {\small\textsf{Sample}} takes a distribution to sample from, whereas {\small\textsf{Observe}} takes a distribution and an observed value. The distribution handler is then given by:
\begin{lstlisting}[]
  handle|$_\textsf{Dist}$||\,|::|\,|(Member Sample es, Member Observe es) |$\Rightarrow$| Prog (Dist |\!|:|\,|es) a |$\rightarrow$| Prog|\,|es|\,|a
  handle|$_\textsf{Dist}$| (Op op k) = case discharge op of
    Right d |$\rightarrow$| case getObs d of Just v |\;$\rightarrow$| (do x |$\leftarrow$| call (|Observe| d v)
                                                |\!|handle|$_\textsf{Dist}$| (k x))
                                Nothing |$\rightarrow$| (do x |$\leftarrow$| call (|Sample| d)
                                                |\!|handle|$_\textsf{Dist}$| (k x)) |\vspace{0.25cm}|
  getObs :: Dist a |$\rightarrow$| Maybe a
\end{lstlisting}

\noindent
The accessor function \lstinline{getObs} retrieves the optional observation associated with a primitive distribution (\secref{distribution-effects}). On encountering a distribution \lstinline{d}, the handler uses \code{getObs} to try to retrieve its observed value; if there is such a value \lstinline{v}, we call a corresponding \code{Observe} operation, and otherwise we call \code{Sample}.

\vspace{-0.1cm}
\subsection{Specialising Multimodal Models}
\label{sec:composing-core-handlers}

Together, these two handlers specialise a multimodal model into the conditional form determined by a particular model environment. Their net effect on the type of the model is illustrated by the following composite handler:
\begin{lstlisting}
  handle|$_\textsf{core}$| :: (Member Observe es, Member Sample es)
            |$\Rightarrow$| Env env |$\rightarrow$| Model env (ObsReader env : Dist : es) a |$\rightarrow$| Prog es a
  handle|$_\textsf{core}$| env = handle|$_\textsf{Dist}$| |$\concat$| (handle|$_\textsf{Read}$| env) |$\concat$| runModel
\end{lstlisting}

\noindent
This handles both \lstinline{Dist} and \lstinline{ObsReader}, replacing all primitive distributions with explicit calls to \code{Sample} and \code{Observe}. As an example, consider the \code{coinFlip} model presented earlier in \figref{coinflip-tree}. Initially applying \code{handle$_\textsf{Read}$} using the example environment \lstinline{((}\obsvar{p} {\small$\assign$}\lstinline{[0.5])}\! $\cons$ \lstinline{(}\obsvar{y} {\small $\assign$}\lstinline{[])} $\cons$ \lstinline{nil)} would produce \code{coinFlip$'$} in \figref{coinflip-tree-env} as an intermediate program, in which all distributions calls have been parameterised by either a concrete observed value or \code{Nothing}; the updated environment would be \lstinline{((}{\small \obsvar{p} $\assign$}\lstinline{[])} \!$\cons$ \lstinline{(}{\obsvar{y} $\assign$}\lstinline{[])} $\cons$ \lstinline{nil)} where \obsvar{p} is fully consumed and \obsvar{y} is unchanged. Then applying \code{handle$_\textsf{Dist}$} to \code{coinFlip$'$} would yield \figref{coinflip-tree-dist}; observed values in primitive distributions are retained, although they are rendered redundant by the information in the \code{Observe} constructor.

\begin{figure}[H]
\vspace{-0.1cm}
\hspace{-0.7cm}
\begin{subfigure}{0.49\textwidth}
  \begin{minipage}{\linewidth}
  \centering\captionsetup[subfigure]{justification=centering}
  \begin{tikzpicture}
	\begin{pgfonlayer}{nodelayer}
		\node [style=none] (3) at (0, 0.75) {\small \textsf{handle}$_{\textsf{Read}}$ \, \textsf{((}{\color{string}\textsf{\#}\textsf{p}} $\assign$\textsf{[0.5])} $\cons$ \textsf{(}{\color{string}\textsf{\#}\textsf{y}} $\assign$\textsf{[])} $\cons$ \textsf{nil)} \, \code{coinFlip}};
		\node [style=none] (4) at (0, -0.25) {$\Downarrow$};
	\end{pgfonlayer}
\end{tikzpicture}
  \end{minipage}
\end{subfigure}
\begin{subfigure}{0.45\textwidth}
  \begin{minipage}{\linewidth}
  \centering\captionsetup[subfigure]{justification=centering}
  \begin{tikzpicture}
	\begin{pgfonlayer}{nodelayer}
		\node [style=none] (0) at (0, 0.75) {\small \textsf{handle}$_{\textsf{Dist}}$\, \code{coinFlip$'$}};
		\node [style=none] (1) at (0, -0.25) {$\Downarrow$};
	\end{pgfonlayer}
\end{tikzpicture}
  \end{minipage}
\end{subfigure}\\
\hspace{-0.8cm}
\begin{subfigure}{0.45\textwidth}
\begin{lstlisting}
    coinFlip|\!$'$| = do
      p |$\leftarrow$| call (Uniform 0 |\!|1 (|\!|Just |\!\!|0.5))
      y |$\leftarrow$| call (|\!|Bernoulli |\!\!|p Nothing)
      return y
\end{lstlisting}
\end{subfigure}
\hspace{0.53cm}
\hspace{-0.2cm}
\begin{subfigure}{0.45\textwidth}
  \begin{lstlisting}
coinFlip|\!$''$| = do
  p |$\leftarrow$| call (|Observe| |\!|(Uniform 0 1 (|\!|Just |\!\!|0.5)) |\!\!|0.5)
  y |$\leftarrow$| call (|Sample| ||(|\!|Bernoulli |\!|p Nothing))
  return y
  \end{lstlisting}
  % \begin{minipage}{\linewidth}
  % \centering\captionsetup[subfigure]{justification=centering}
  % \tikzfig{figures/tree-3}
  % \end{minipage}
\end{subfigure}\\
\hspace{-1cm}
\hspace{0.9cm}
\begin{subfigure}{0.4\textwidth}
  \vspace{-0.25cm}
  \caption{Reading from the model environment}
  \label{fig:coinflip-tree-env}
\end{subfigure}
\hspace{0.6cm}
\begin{subfigure}{0.48\textwidth}
  \vspace{-0.25cm}
  \caption{Distributions interpreted to sample or observe}
  \label{fig:coinflip-tree-dist}
\end{subfigure}
\vspace{-0.2cm}
\caption{\code{coinFlip} model: handling effects}
\label{fig:coinflip-handled-tree}
\vspace{-0.3cm}
\end{figure}

\noindent
This initial pipeline of effect handling turns a multimodal model into a form suitable for further specialisation by a simulation or inference algorithm, which we show in \secref{simulation-inference}.

\subsection{Extending Models with Additional Effects}
\label{sec:extending-models-with-effects}

When building models, users are not restricted to using only the two base effects \lstinline{Dist} and \lstinline{ObsReader}: the effect signature \code{es} in \lstinline{Model env es a} can be easily extended with an arbitrary desired effect \lstinline{e}, by first constraining the model with \lstinline{Member e es}, and then handling \code{e} with a corresponding handler.

As an example, we revisit the SIR model in \secref{the-sir-model} where \figref{sir} plotted \emph{all} intermediate \LatType\ values of susceptible, infected, and recovered individuals over $n$ days, despite our implementation \code{hmm$_{sir}$} only returning the final one:
\begin{lstlisting}
  hmm|$_{sir}$| :: (Observables env '[|\color{string}{``$\xi$''}|] Int, Observables env '[|\color{string}{``$\beta$''}|, |\color{string}{``$\gamma$''}|, |\color{string}{``$\rho$''}|] Double)
        |$\Rightarrow$| Int |$\rightarrow$| |\LatType{}| |$\rightarrow$| Model env es |\LatType{}|
  hmm|$_{sir}$| |$n$| = hmm transPrior|$_{sir}$| obsPrior|$_{sir}$| trans|$_{sir}$| obs|$_{sir}$| |$n$|
\end{lstlisting}

To record all $sir$ values produced by \code{hmm$_{sir}$}, we introduce the constraint \lstinline{Member (Writer w) es} to require \lstinline{es} to contain the well-known effect \lstinline{Writer w} (omitted), representing computations that produce a stream of data of type \lstinline{w}; here we choose \lstinline{w} to be a list of $sir$ values \code{[\LatType{}]}. The transition model \code{trans$_{sir}$} can then use the \lstinline{Writer} operation \code{tell} to concatenate each new $sir$ value to the existing trace of values:

\begin{lstlisting}
  |\!|tell     |\!|:: Member (Writer w) es |$\Rightarrow$| w |$\rightarrow$| Model env es () |\vspace{0.1cm}|
  trans|$_{sir}$| :: Member (Writer [|\LatType{}|]) es |$\Rightarrow$| TransModel env es TransParams |\LatType{}|
  trans|$_{sir}$| (|TransParams| |$\beta$| |$\gamma$|) |$sir$| = do
    |$sir'$| |$\leftarrow$| (trans|$_{si}$| |$\beta$| |\texttt{>=>}| trans|$_{ir}$| |$\gamma$|) |$sir$|
    |\!\!|tell [|$sir'$|]
    return |$sir'$|
\end{lstlisting}
\noindent

Models with user-specified effects can then be easily reduced into a form suitable for specialisation under a model environment (\secref{composing-core-handlers}), by handling those effects beforehand with a suitable handler:
\begin{lstlisting}
  handle|$_{\textsf{Writer}}$| :: Monoid w |$\Rightarrow$| Model env (Writer w : es) a |$\rightarrow$| Model env es (a, w) |\vspace{0.15cm}|
  hmm|$_{sir}'$| :: (Observables env '[|\color{string}{``$\xi$''}|] Int, Observables env '[|\color{string}{``$\beta$''}|, |\color{string}{``$\gamma$''}|, |\color{string}{``$\rho$''}|] Double)
          |$\Rightarrow$| Int |$\rightarrow$| |\LatType{}| |$\rightarrow$| Model env es (|\LatType{}|, [|\LatType{}|])
  hmm|$_{sir}'$| |$n$| = handle|$_{\textsf{Writer}}$| |\concat| hmm|$_{sir}$| |$n$|
\end{lstlisting}
Here, composing \code{hmm$_{sir}$} with \code{handle$_\textsf{Writer}$} interprets the \code{tell} operations arising from the transition model, producing a new SIR model \code{hmm$_{sir}'$} that returns the trace of \code{$sir$} values as an additional output of type \code{[\LatType{}]}.
\section{Simulation and Inference as Effect Handlers}
\label{sec:simulation-inference}

Simulation or inference over a model is expressed through the semantics we assign to its \lstinline{Sample} and \lstinline{Observe} effects; in principle we could therefore define model execution in terms of two handlers, one for each of those effects. However, many algorithmic approaches such as Monte-Carlo methods \cite{chong2010monte} often rely on common mechanisms, such as sample tracing, probability mapping, and model reparameterisation \cite{robert1998reparameterization}. By identifying some of these mechanisms as basic building blocks, the possibility arises of composing them in novel ways, giving rise to many useful variants of algorithms \cite{mbayes}. This motivates a compositional approach in which aspects of model execution can be defined and extended modularly.

In this section, we use effect handlers to define a series of composable program transformations for probabilistic programs, which iteratively refines a model by installing new effects around existing ones. The result is an interpretation of the model in the context of specific algorithms for simulation in \secref{simulation-as-effect-handlers} and inference in \secref{inference-as-effect-handlers}.

\vspace{-0.1cm}
\subsection{Simulation as Effect Handlers}
\label{sec:simulation-as-effect-handlers}
Simulation can be considered the most basic form of model execution. It runs the provided model as a generative process, using observed data from the model environment when available and otherwise drawing new samples. At the end, it returns a model output, plus a \textit{sample trace} uniquely identifying each \emph{runtime} \code{Sample} operation with its sampled value; this becomes especially pertinent (\secref{mh}) for the correctness and implementation of generic inference algorithms \cite{tolpin2016design}.

To support sample traces, each runtime \code{Sample} occurrence thus needs a unique dynamic address $\alpha$ assigned; in our actual embedding, this feature is implemented by \lstinline{handle}$_\textsf{Dist}$. For simplicity, we omit these low-level details, and assume operations {\small\textsf{Sample}} and \code{Observe} are now parameterised by an address {\small $\alpha$} of abstract type \lstinline{Addr}. The type of sample traces, \lstinline{STrace}, is then a map from addresses \code{$\alpha$} to values of abstract type \lstinline{PrimVal} that primitive distributions can generate (concretely, \lstinline{PrimVal} is an open sum, but one can avoid this using dependent maps):

\begin{lstlisting}
  type STrace = Map Addr PrimVal |\vspace{0.1cm}|
\end{lstlisting}
Updating this sample trace is to be performed by the well-known \lstinline{State} effect:

\begin{lstlisting}
  data State s a where
    Modify :: (s |$\rightarrow$| s) |$\rightarrow$| State s () |\vspace{0.15cm}|
  handle|$_\textsf{State}$| :: s |$\rightarrow$| Prog (State s : es) a |$\rightarrow$| Prog es (a, s)
\end{lstlisting}
\noindent
Its operation \code{Modify} takes a function of type \code{s $\rightarrow$ s} and applies this to state \code{s}. Its handler \code{handle$_\textsf{State}$}, given an initial state, additionally returns the final state that results from handling a program.

We implement the tracing of samples as the \emph{program transformation} \lstinline{traceSamples} below, which is a handler that installs a runtime \lstinline{State STrace} effect after each \code{Sample} operation:

\begin{lstlisting}[escapechar=|]
  traceSamples :: (Member Sample es, Member (State STrace) es) |$\Rightarrow$| Prog es a |$\rightarrow$| Prog es a
  traceSamples (Op op k) = case prj op of
    Just (|Sample| d |$\alpha$|) |$\rightarrow$| |Op| op (|$\lambda$|x -> do |\,|call (Modify (Map.insert |$\alpha$| x))
                                            traceSamples (k x))
    Nothing            -> Op op (traceSamples . k)
\end{lstlisting}
Notice that we apply \lstinline{prj} (\secref{effect-infrastructure}) to \lstinline{op}, rather than \lstinline{discharge}; this is because we do not intend to handle any effects, and so we place no constraints on the order of the effect signature \lstinline{es}. Rather, we leave any \code{Sample} operations unhandled as \lstinline{op} in the program, and construct a new continuation: this takes the future output \lstinline{x} from \code{Sample} and calls a \lstinline{Modify} operation to store \lstinline{x} at address \code{$\alpha$} in the sample trace, before continuing with the original continuation \lstinline{k}. The case of \code{prj} producing \code{Nothing} follows the same pattern as \code{discharge} producing \code{Left op'}.

We can now define the handlers for \lstinline{Observe} and \lstinline{Sample} orthogonally from this transformation. For \code{Observe} operations, \lstinline{handle}$_\textsf{Obs}$ performs no conditioning side-effects, and simply needs to return the observed value \lstinline{y} to its continuation \lstinline{k}:

\begin{lstlisting}
  |handle$_\textsf{Obs}$| :: Prog (Observe : es) a |$\rightarrow$| Prog es a
  |handle$_\textsf{Obs}$| (Op op k) = case discharge op of
    Right (|Observe| d y _) |$\rightarrow$| handle|$_\textsf{Obs}$| (k y)
\end{lstlisting}

For \code{Sample} operations, \lstinline{handle}$_\textsf{Samp}$ takes the provided distribution \lstinline{d} and applies the function \lstinline{sampleIO}, defined using an external statistics library; the generated value \lstinline{v} is then passed to \lstinline{k}:
% In our actual implementation, \lstinline{Addr} values are integer-indexed observable variable names e.g. \lstinline{(}{\color{string}\lstinline{#y}}{, 0)}, and \lstinline{PrimVal} is an open sum of primitive types e.g. \lstinline{Int} and \lstinline{Bool}.
\begin{lstlisting}
  handle|$_\textsf{Samp}$| :: Prog '[|\!|Sample] a |$\rightarrow$| IO (a, STrace)
  |handle$_\textsf{Samp}$| (Op op k) = case discharge op of
    Right (|Sample| d _) |$\rightarrow$| do v |$\leftarrow$| sampleIO d
                             |\,|handle|$_\textsf{Samp}$| (k v)|\vspace{0.1cm}|
  sampleIO :: Dist a |$\rightarrow$| IO a
\end{lstlisting}
\noindent
Running this dispatches the final effect in \lstinline{Prog} to produce an \lstinline{IO} effect, hence it is always executed as the last handler where only \code{Sample} operations can occur in the program.

The complete definition for simulation is then given by the handler composition \lstinline{runSimulate}:

\begin{lstlisting}
  runSimulate :: es |$\sim$| '[|\!|ObsReader env, Dist, State STrace, Observe, Sample]
              |$\Rightarrow$| Env env |$\rightarrow$| Model env es a |$\rightarrow$| IO (a, STrace)
  runSimulate env = handle|$_\textsf{Samp}$| |$\concat$| handle|$_\textsf{Obs}$| |$\concat$| (handle|$_\textsf{State}$| Map.empty) |$\,\concat\,$| traceSamples |$\,\concat$| (handle|$_\textsf{core}$| env)
\end{lstlisting}
\noindent
Above depicts how the logic of model execution can be decomposed into a modular and transparent system. The concrete effects in \lstinline{es}, specified by the type coercion \code{($\sim$)}, are kept abstract to the user interested in simply building and using models; simultaneously, the compositional nature of handlers makes it easy for programmers who wish to implement and extend new forms of model execution, as demonstrated next in \secref{inference-as-effect-handlers}.

As a last remark, the reader may notice that the top-level function \lstinline{simulate} (as seen in \secref{the-sir-model}), which uses \lstinline{runSimulate}, will differ slightly in its type signature, which we give now:
\begin{lstlisting}
  simulate :: es |$\sim$| '[|\!|ObsReader env, Dist, State STrace, Observe, Sample]
          |$\Rightarrow$| (x |$\rightarrow$| Model env es a) |$\rightarrow$| Env env |$\rightarrow$| x |$\rightarrow$| IO (a, Env env)
\end{lstlisting}
This allows for better composition when simulating a model over many inputs \code{x}. Then for type-safe user-access to an \lstinline{STrace} structure, this is reified into an \textit{output} environment of type \lstinline{Env env}; for this, we use simple type-level programming to extract values from the trace whose addresses $\alpha$ (in the full implementation) are indexed by an observable variable name from the input environment \code{env}. A similar approach is taken for Likelihood Weighting (\code{lw}) and Metropolis Hastings (\code{mh}).

% Simulating many times over multiple model inputs is then a matter of applying the functional combinators \lstinline{map} and \!\lstinline{replicate}\!.

% \todo{Do we include the concrete definition of simulate?}

% \begin{lstlisting}
%   simulate n model xs env = concat <$> mapM (replicateM n |$\concat$| handle|$_\textsf{simulate}$| env |$\concat$| model) xs
% \end{lstlisting}
% The definition of \lstinline{simulate} says: we perform \lstinline{n} simulations (via \lstinline{replicateM}) of \lstinline{model} under the environment \lstinline{env} for each model input in \lstinline{xs} (via \lstinline{mapM}), the result of which is flattened (via \lstinline{concat}).

\subsection{Inference as Effect Handlers}
\label{sec:inference-as-effect-handlers}

Approximative Bayesian inference attempts to learn the posterior distribution of a model's parameters given some observed data. We reuse the ideas introduced in \secref{simulation-as-effect-handlers} to implement Likelihood Weighting and Metropolis Hastings as inference algorithms.

\subsubsection{Likelihood Weighting (LW)} If one uses simulation as a process for randomly proposing model parameters, the LW algorithm \cite{van2018introduction} then assigns these proposals a \textit{weight}, that is, the total likelihood of them having generated some specified observed data.

Its implementation is in fact extremely simple, as most of the work has been done when implementing \lstinline{runSimulate}. The only change is to how the \lstinline{Observe} effect is interpreted:

\begin{lstlisting}
  handle|$_\textsf{ObsLW}$| :: Double |$\rightarrow$| Prog (Observe : es) a |$\rightarrow$| Prog es (a, Double)
  |handle$_\textsf{ObsLW}$| lp (Op op k) = case discharge op of
    Right (|Observe| d y _) -> handle|$_\textsf{ObsLW}$| (lp + logProb d y) (k y)
  |handle$_\textsf{ObsLW}$| lp (Val x)   = return (x, lp) |\vspace{0.15cm}|
  logProb :: Dist a |$\rightarrow$| a |$\rightarrow$| Double
\end{lstlisting}
The function \lstinline{handle}$_\textsf{ObsLW}$ is now parameterised by a log probability \lstinline{lp}, where at each operation \code{Observe d y $\alpha$}, it computes and adds to \lstinline{lp} the log probability of distribution \lstinline{d} having generated observed value \lstinline{y}. The total log probability is returned upon reaching \code{Val x}.

The complete definition for a single iteration of LW is given as \lstinline{runLW}:

\begin{lstlisting}
  runLW :: es |$\sim$| '[|\!|ObsReader env, Dist, State STrace, Observe, Sample]
        |$\Rightarrow$| Env env |$\rightarrow$| Model env es a |$\rightarrow$| IO ((a, STrace), Double)
  runLW env = handle|$_\textsf{Samp}$| |$\concat$| (handle|$_\textsf{ObsLW}$| 0) |$\concat$| (handle|$_\textsf{State}$| Map.empty) |$\concat$| traceSamples |$\concat$| (handle|$_\textsf{core}$| env)
\end{lstlisting}
This is identical to simulation but we now use \lstinline{handle}$_\textsf{ObsLW}$ instead of \lstinline{handle}$_\textsf{Obs}$; running this returns the log-likelihood of the values in \lstinline{STrace} giving rise to the observed data in provided environment \lstinline{env}. Similar to \lstinline{runSimulate} and \lstinline{simulate}, \lstinline{runLW} is called via a top-level function \lstinline{lw} (omitted), but is instead performed iteratively to produce a trace of weighted parameter proposals (\figref{example-lin-regr-inf}).

% \begin{lstlisting}
%   lw :: es |$\sim$| '[|\!|ObsReader env, Dist, State STrace, Observe, Sample]
%      |$\Rightarrow$| Int |$\rightarrow$| (x |$\rightarrow$| Model env es a) |$\rightarrow$| (x, Env env) |$\rightarrow$| IO [(Env env, Double)]
% \end{lstlisting}
% This iterates \lstinline{runLW}, resulting in a list of weighted parameter proposals stored in output environments -- an example of this was visualised in \figref{example-lin-regr-inf}.

Likelihood Weighting, however, becomes ineffective as the number of random variables sampled from increases: as values are freshly generated for \emph{all} \code{Sample} operations, achieving a high likelihood means sampling an entire set of likely proposals. Our next, final example offers a solution to this:

\subsubsection{Metropolis Hastings (MH)}
\label{sec:mh}
MH \cite{wingate2011lightweight} supports \emph{incremental} parameter proposals by randomly choosing a ``proposal address'' $\alpha_0$  at the start of each iteration; this denotes the address from which a new sample is to be drawn, where all other addresses are to instead reuse old samples from previous iterations. At the end of an iteration, we decide whether to accept the new sample by comparing the \textit{individual} log probabilities of each probabilistic operation.

To support this, we take a similar approach to the one shown for \code{traceSamples}: we define the type \lstinline{LPTrace} to map addresses of probabilistic operations to their log probabilities, and then the program transformation \lstinline{traceLPs} to update this as a state.
\begin{lstlisting}
  type LPTrace = Map Addr Double |\vspace{0.1cm}|
  traceLPs ::|\,|(Member Observe es, Member (State LPTrace) es) |$\Rightarrow$| Prog es a |$\rightarrow$| Prog es a
  traceLPs (Op op k) = case prj op of
    Just (|Observe| d y |$\alpha$|) -> |Op| op (|$\lambda$|x -> do call (Modify (Map.insert |$\alpha$| (logProb d y|\!|)))
                                             traceLP (k x)
\end{lstlisting}
On matching against \code{Observe d y $\alpha$}, the above installs a \lstinline{State} \lstinline{LPTrace} effect by defining a new continuation that stores the log probability of \lstinline{d} having generated \lstinline{y}. The same is done for \code{Sample} operations, but we compute the log probability of the values sampled.

All of the conditioning is in fact taken care of by \code{traceLPs}, so the handler for \code{Observe} operations only needs to return any observed values to their continuations -- its implementation is therefore the same as for simulation (\secref{simulation-as-effect-handlers}). What is left to define is how \code{Sample} is interpreted:

\begin{lstlisting}
  handle|$_\textsf{SampMH}$| :: STrace |$\rightarrow$| Addr |$\rightarrow$| Prog '[|\!|Sample] a |$\rightarrow$| IO a
  |handle$_\textsf{SampMH}$ sTrace $\alpha_0$ (Op op k)| = case discharge op of
      Right (|Sample d $\alpha$|) -> do x |$\leftarrow$| |lookupSample sTrace d $\alpha$ $\alpha_0$|
                                |handle$_\textsf{SampMH}$ $\alpha_0$ sTrace (k x) \vspace{0.2cm}|
  lookupSample :: STrace |$\rightarrow$| Dist a |$\rightarrow$| Addr |$\rightarrow$| Addr |$\rightarrow$| IO a
\end{lstlisting}

\noindent
The handler \code{handle$_\textsf{SampMH}$} takes the sample trace \lstinline{sTrace} of the previous MH iteration, and an address $\alpha_0$ denoting the proposed sample site. On matching against an operation \code{Sample d $\alpha$}, the function \lstinline{lookupSample} will generate a new sample if $\alpha$ matches $\alpha_0$, or if no previous MH iterations have yet generated a sample for it; otherwise, it reuses the old sample value of $\alpha$ found in \lstinline{sTrace}.

The final definition for a single iteration of MH is given as \code{runMH}:

\begin{lstlisting}
  runMH :: es |$\sim$| '[|\!|ObsReader env, Dist, State STrace, State LPTrace, Observe, Sample]
        |$\Rightarrow$| Env env |$\rightarrow$| STrace |$\rightarrow$| Addr |$\rightarrow$| Model env es a |$\rightarrow$| IO ((|\!|a, STrace|\!|), LPTrace)
  runMH env sTrace |$\alpha_0$| = (handle|$_\textsf{SampMH}$| sTrace |$\alpha_0$|) |$\concat$| handle|$_\textsf{Obs}$|
                      |$\concat$| |\,|(handle|$_\textsf{State}$| Map.empty) |\,||$\concat$| |\,|(handle|$_\textsf{State}$| Map.empty)
                      |$\concat$|  traceLPs |$\concat$| traceSamples |$\concat$| (handle|$_\textsf{core}$| env)
\end{lstlisting}

\noindent
This reuses many of the building blocks of simulation, and extends these with the transformation \lstinline{traceLPs} and handler for \lstinline{State} \lstinline{LPTrace}; the \code{handle$_\textsf{SampMH}$} variant is then used instead of \code{handle$_\textsf{Samp}$}.

Although \code{runMH} adheres to the MH algorithm in terms of how it samples and conditions, it does not decide the proposal address $\alpha_0$ at the start of an MH iteration, nor whether newly proposed parameters are accepted at the end of an iteration. We keep this logic distinct from effect handlers, and instead implement it in a wrapper function \lstinline{mh} (omitted) which folds over iterations of \code{runMH}, propagating information from the previous MH iteration to support decision making in the next; the end result generates a trace of accepted parameter proposals (\figref{sir-inf}).
% , and then (for the purposes of this paper) concatenates all sample traces into a single output model environment. The result generates a trace of accepted parameter proposals, as was shown in \figref{sir-inf}.
% \begin{lstlisting}
%   mh :: es |$\sim$| '[|\!|ObsReader env, Dist, State STrace, State LPTrace, Observe, Sample]
%      |$\Rightarrow$| Int |$\rightarrow$| (x |$\rightarrow$| Model env es a) |$\rightarrow$| (x, Env env) |$\rightarrow$| IO (Env env)
% \end{lstlisting}
% This uses \code{runMH} to perform MH iterations, and then (for the purposes of this paper) concatenates all sample traces into a single output model environment. The result generates a trace of accepted parameter proposals, as was shown in \figref{sir-inf}.
\section{Evaluation}
\label{sec:evaluation}

\subsection{Quantitative Evaluation}

We compare our language's performance, which for simplicity is implemented using \code{freer-simple}\footnote{\url{https://github.com/lexi-lambda/freer-simple}\unskip}, against two state-of-the-art PPLs. First, MonadBayes \cite{mbayes}, a Haskell embedded language which uses a monad transformer library (\code{mtl}\footnote{\url{https://github.com/haskell/mtl}}) approach as an effect system for PPLs, but does not support multimodal models. Second, Turing \cite{turingjl}, which achieves multimodal models via macro compilation in Julia as a host language.

Our evaluation strategy uses a set of popular benchmarks \cite{ppl-benchmarking}, comparing simulation (SIM), Likelihood Weighting ({LW}), and Metropolis Hastings ({MH}) as forms of execution algorithms. We apply these to the following example models: linear regression, hidden Markov model, and latent dirichlet allocation. These benchmarks, given fully in \apdxref{benchmarks}, are performed on an AMD Ryzen 5 1600 Six-Core Processor with 16GB of RAM.

Across all models, our performance scales linearly with the number of samples each algorithm generates; this remains mostly the case when varying the dataset size to models, except for inference on hidden Markov model where we begin to scale quadratically.

Against Turing, we are on average \code{4.0x} and \code{1.2x} faster for SIM and LW. As our benchmarks do not consider the overhead from Turing's macro-compilation stage for specialising multimodal models, this may indicate our runtime approach to model specialisation does not notably affect performance. MonadBayes is on average \code{3.3x} faster than us for SIM and has close to constant-time performance for \code{LW}, making it difficult to compare the latter. In contrast to us, they choose to not store and update sample traces for SIM and LW; this effectively means their performance is impacted only by the cost of sampling operations (of which there are naturally fewer for LW), whereas we incur overhead from additional \lstinline{State} operations. Another factor we heed is the performative difference between \code{freer-simple} and \code{mtl} for large, non-synthetic programs, which still requires investigation.

For MH in particular, Turing and MonadBayes are on average \code{3.1x} and \code{1.8x} faster. Our implementation is naive in that we traverse the whole sample trace for MH updates despite a fixed number of variables being updated -- this is straightforward to amend \cite{wingate2011lightweight}. We also remark that Turing and MonadBayes only perform MH updates to \emph{one} variable per iteration, and do not consider the dependent variables which consequently need updating \cite{kiselyov2016problems}; our implementation does account for this and hence incurs costs for updating groups of dependencies, which can be larger for models with more complex dependency graphs.

% We believe part of this is due to \emph{all} of our handlers being applied each model execution, whereas we suspect partially pre-evaluated models can instead be executed by reconsidering the order of handling.

We find our language competitively performant despite not yet having explored potential optimisation, and expect to benefit greatly from off-the-shelf Haskell techniques such as inlining and more efficient data structures. Exploring performance of effect handlers with PPLs in general is an important and challenging topic we plan to investigate separately, including techniques specific to algebraic effects, such as the codensity monad \cite{voigtlander2008asymptotic}, and alternative representations of effectful programs \cite{wu2015fusion, polysemy}. We also observe that \emph{all} of our handlers are applied on each model execution, whereas we suspect partially pre-evaluated models can be executed by reconsidering the order of handling; this may lead to substantial performance gains.

% We note that the cost of inference is typically less than that of simulation as one generally provides more observed data to the model for inference, resulting in less sampling and hence less computational overhead from pseudo-random number generators -- we therefore suspect the slopes (performance) with respect to inference are more attributed to algorithmic overhead and language abstractions used in the system design.

% We generally expect monadbayes to be more performant given its model implementations are specific to simulation or inference
%

\subsection{Qualitative Evaluation}
\label{sec:qualitative-eval}
% We have found the ability to represent models as syntax trees as a highly desirable property for model-based PPLs which algebraic effects accommodates for: this allows for a very flexible interpretation of models and probabilistic operations by inference algorithms, and being able to performing passes of transformations over these trees breaks lets us incrementally implement the semantics of inference in a fine-grained manner. Additionally, the algebraic effect approach is sufficiently minimalistic in capturing only the effects we care about as syntax, whilst letting us freely leverage existing host language operations alongside them.

\begin{table}
  \vspace{-0.2cm}
  \small
  \caption{Comparison of PPLs in terms of supported features for models, where \fullcirc[0.8ex] is full support, \halfcirc[0.8ex] is partial support, and \emptycirc[0.8ex] is no support. Modular models are those that can be defined in terms of other models. }
  \vspace{-0.2cm}
  {\setlength{\extrarowheight}{2pt}
  \begin{tabular}{|l|c|c|c|c|c|c|c|c|c|}
  \hline
  Supported model features & \!\!  ProbFX \!\!  & \! Gen \! & \!Turing\!     & \!Stan\!    & \!Pyro\!      &   \!MonadBayes\! & \!Anglican\!   & \!WebPPL\!      \\ \hline
  Multimodal        &   \fullcirc[0.8ex]     &  \fullcirc[0.8ex]  & \fullcirc[0.8ex]  & \fullcirc[0.8ex]   &  \halfcirc[0.8ex] &   \emptycirc[0.8ex]           & \emptycirc[0.8ex]           &  \emptycirc[0.8ex]            \\ \hline
  Modular            &   \fullcirc[0.8ex]    &  \fullcirc[0.8ex]  & \fullcirc[0.8ex]  &   \emptycirc[0.8ex]        &  \fullcirc[0.8ex]   & \fullcirc[0.8ex]     & \fullcirc[0.8ex]  & \fullcirc[0.8ex] \\ \hline
  Higher-order       &  \fullcirc[0.8ex]     &  \halfcirc[0.8ex]  & \emptycirc[0.8ex]           &  \emptycirc[0.8ex]         & \fullcirc[0.8ex]    &  \fullcirc[0.8ex]    & \halfcirc[0.8ex]    &  \fullcirc[0.8ex] \\ \hline
  Type-safe          &  \fullcirc[0.8ex]      &  \emptycirc[0.8ex]  & \emptycirc[0.8ex]          &  \fullcirc[0.8ex]  &  \emptycirc[0.8ex]         &  \fullcirc[0.8ex]   & \emptycirc[0.8ex]    & \emptycirc[0.8ex] \\ \hline
  \end{tabular}
  }
  \label{fig:feature-comparison}
  \vspace{-0.15cm}
\end{table}

Finally, we compare our supported features across a larger range of modern PPLs in \figref{feature-comparison}. To the best of our knowledge, our language is the first to fully support both multimodal and higher-order models. Models in Gen \cite{genjl}, denoted with the special \lstinline{@gen} syntax, have support for higher-order interactions with other \lstinline{@gen} terms; however, applying a standard higher-order function to a model, at least without programmer intervention, will escape the tracing of the model's computation. In Pyro \cite{pyro}, models are Python functions and as such are first-class, but support for multimodal models is limited: random variables may have only \emph{one} observed value at runtime, and so Pyro expects this value to be a matrix where all contained data is conditioned on simultaneously. This approach cannot be used for models with sequential behaviour such as HMMs.

As far as we know, our language is also the first general-purpose PPL with multimodal models in a statically typed paradigm. Stan \cite{stan} is type-safe but special-purpose. Other general-purpose PPLs with multimodality are dynamically typed (Pyro) or use just-in-time compilation (Turing, Gen), and so do not guarantee that the observed values assigned to random variables are correctly typed, or that these variables exist; however, some progress has been made towards a type-safe version of Gen \cite{lew2019trace}. Although MonadBayes is statically typed, we are not aware of any attempts to extend its \code{mtl} approach to support multimodal models.

We note that as our language is experimental, the range of model execution algorithms we present so far is small. Also, whilst we provide basic utilities such as IO debugging and type-safe interfacing with sample traces, richer PPL features such as run-time inference diagnostics have not yet been implemented. We believe our language's infrastructure, having embedded into an algebraic effect setting, can readily support extensions in both of these areas.
% \newpage
\section{Conclusion}
\label{sec:conclusion}

Probabilistic programming is an active research topic, with applications ranging from artificial intelligence \cite{tran2017deep} to market research \cite{letham2016bayesian}. Many existing PPLs are too low-level to capture multimodal descriptions of models, or too rigid to allow models to be easily reused. In this paper we used an algebraic effects embedding to implement a PPL where multimodal models are first-class citizens that can be modularly defined and easily combined.

\subsection{Related Work}
\label{sec:related-work}
% w.r.t argument order & prog being in a monad

\subsubsection{Effect Abstractions for PPLs}

Although using interpreters to execute models is a common design pattern in PPLs (e.g.~\citet{tolpin2016design, stan, pymc3}), viewing these interpreters formally as algebraic effect handlers has little precedent in the literature. \citet{scibior2015effects} first demonstrate in Haskell how rejection sampling can be abstracted into handlers; \citet{moore2018effect} use Python \textit{context managers} for sample tracing and conditioning. Although this latter approach takes inspiration from effect handlers, the details are quite different: operations are methods rather than uninterpreted syntax, handlers are coroutines which programs explicitly invoke, and there is no type discipline associating operations and handlers with specific effects.

A notable alternative effect system is the \emph{monad transformer library} (\code{mtl}) \cite{liang1995monad} approach adopted in MonadBayes \cite{mbayes}. The observation behind MonadBayes is that inference algorithms can be decomposed into monadic building blocks which can then be assembled into different stacks of monad transformers. This resembles our approach in using program transformations to augment models with additional inference-specific behaviours, but rather than successively refining a syntactic representation of the model by composing handlers, in MonadBayes one constructs a directly executable model by composing monadic functions. (A simplifying assumption of our approach is that handlers are always pure, producing intermediate programs of type \lstinline{Prog es a}; permitting handlers to return the transformed program in a monad requires additional infrastructure to forward the monadic effect \cite{schrijvers2019monad}.)

% However, not all computational effects behave in this way, and it is conceivable that users may want to interact with effects that instead return \lstinline{m (Prog es a)} for some monad \lstinline{m}.

In terms of usability, at least for model construction, we find the \code{mtl} approach of MonadBayes offers a similar experience to algebraic effects: in particular both support abstract effect signatures for model components, so that programs can be built ``bottom-up'' with effect constraints propagated upwards. For implementing sophisticated inference algorithms, monad transformers can be tricky. The standard \code{mtl} design pattern allows the structure of a transformer stack to remain abstract, and operations belonging to a particular monad \code{m} to be invoked at any point lower in the stack by having monads below it relay the operations up to \code{m}. In MonadBayes, however, several monads perform their own \code{sample}/\code{observe} side-effects, and may choose not to relay these operations up the stack. The network of relaying is non-trivial and carried out implicitly via type class instances, making the final behaviour rather opaque. By contrast, such logic can be more transparently expressed as a single effect handler. Monad transformers also seem to be less modular for expressing algorithms parameterised by alternative behaviours, e.g.~static vs. dynamic tracing of random choices \cite{mbayes}, requiring a substantial amount of boilerplate for each alternative monad instance.

\vspace{-0.1cm}

\subsubsection{Embedding PPLs}

Tagless-final shallow embedding, as conceived by \citet{kiselyov2012typed}, has also been used as an embedding technique for PPLs \cite{kiselyov2016probabilistic, kiselyov2009embedded, narayanan2016probabilistic}. The syntax of the PPL is captured as a type class, and type class instances used to provide interpretations of programs under particular inference algorithms. This is well suited to mapping programs uniformly into a semantic domain, but we found it difficult to use the approach to iterate transformations over models (such as interpretations to \lstinline{sample} and \lstinline{observe}) in order to implement inference compositionally.

Free monads have also seen use: \citet{scibior2015practical} embed primitive and conditional distributions using an intermediate free monad representation. This bears an initial resemblance to our approach, but the semantics are instead provided using type classes, and their direct encoding of conditional distributions means models are not multimodal. Later work by \citet{mbayes} more closely coincides with our approach; their implementation of Metropolis-Hastings uses free monad transformers \cite{schrijvers2019monad} to encode sampling operations as syntax, allowing models to be executed so that sampling can either invoke an \code{IO} effect or reuse previous samples.

% In terms of PPLs that support multimodal models, most typically require some form of special compilation stage for models. General-purpose PPLs such as Turing.jl and Gen.jl tend to take advantage of the macro-compilation features of their host language to rewrite the syntax trees of models into executable functions \cite{turingjl, genjl}. Special-purpose PPLs, e.g. Stan and BUGS, implement models as their own language construct, which are then compiled to entirely different languages \cite{stan, bugs}.

\subsubsection{Observed Data for Multimodal Models}

For providing observations to multimodal models, Pyro \cite{pyro} and Gen \cite{genjl} take the most similar approach to ours, letting users provide mappings between variable names and observed values to a top-level function representing a model. Pyro wraps the model in a runtime context manager that constrains the values of variables to the provided data, whereas Gen uses the data to macro-expand the model into either a static computation graph or a standard function. To support multiple observed values for the same variable, Pyro relies on this variable to expect a matrix, as mentioned in \secref{qualitative-eval}; Gen better facilitates multiple sequential observations against a single variable, but requires unique names to be provided for each runtime occurrence. Our approach allows a list of observations to be associated with a given variable.

Special-purpose PPLs such as Stan \cite{stan} and WinBUGS \cite{bugs} specify models and observed data separately via bespoke language constructs; common variable names are then resolved and linked during compilation.

Although our specific approach to model environments appears to be novel, work by \citet{lew2019trace} is closely related. They investigate row polymorphism for assigning types to names of random choices when tracing probabilistic computations; we hope to build on this idea when formalising our language, in particular for characterising the execution space of a multimodal model under a fixed model environment.

\subsection{Future Work}
\label{sec:future-work}
% For future work, we recognise two directions. The first is our work-in-progress minimal calculus and semantics of our language, which formalises 1) the necessary language constructs of multimodal models and model environments in an algebraic effect setting, and 2) the correctness of our effect handler approach for inference and specialising multimodal models; the general notions presented in this formalism can then be implemented in languages with sufficient support.

% This also explores possible settings This Here we explore row polymorphism for effect types \cite{leijen2014koka} that is an affine type-system

Our implementation currently responds to running out of observed values in the model environment by switching to sampling, as is typical in PPLs \cite{turingjl,pyro}. However, it may be possible in some settings to use type-level naturals to statically constrain the number of observations required. Alternatively, a dynamic check to signal when too many or too few observations are provided would be easy to implement.

% We also investigate how naming conflicts between observable variables should be resolved when combining models. Constraint kinds are currently used to enforce uniquely named variables in model environments, but an additional method we consider is a renaming mechanism that allows observable variables to be rebound to a global specification of names.

We are also investigating how naming conflicts between observable variables should be resolved when combining models. This is not considered an issue by most existing PPLs; many  require programmers to uniquely name each dynamic random variable instance \cite{genjl, pymc3}, perhaps failing at runtime if this condition is not met \cite{pyro}. Our language is type-safe in allowing model environments to have typed, orthogonally combinable variables (via constraint kinds), but for additional modularity, we are also considering a renaming mechanism for rebinding observable variables when name clashes arise.

A related topic is when the programmer statically refers to the same observable variable more than once in a model:
\begin{lstlisting}
  do x|$_1$| |$\leftarrow$| normal 0 1 |\obsvar{x}|
     x|$_2$| |$\leftarrow$| normal 0 2 |\obsvar{x}|
     return (x|$_1$| + x|$_2$|)
\end{lstlisting}
Technically, this is an invalid use of a random variable, resulting in an ill-formed model where \obsvar{x} is (confusingly) distributed according to two different distributions. For now, programmers must take care not to misuse observable variables in this way; a solution we intend to explore is an affine type system for model contexts which will disallow multiple static uses of the same observable variable. We are also working on a formalisation of our language and embedding technique.

Lastly, we aim to explore how using effect handlers as program transformations can help with the implementation of sophisticated, compositional inference algorithms such as SMC$^2$ \cite{doucet2001sequential} and PMMH \cite{chopin2002sequential}; we anticipate needing to reason about interactions with new effects, for example with non-determinism and exceptions, and whether such effects distribute over \cite{wu2014effect} and commute with each other \cite{gibbons2011just} in a probabilistic setting. For scaling up to advanced inference techniques that require gradient information, e.g. HMC and NUTS \cite{hoffman2014no}, models also need to be differentiable. This is generally done in PPLs via \emph{automatic differentation} (AD), a procedure for interpreting standard programs as differentiable functions. It may be possible to support this using our existing effects infrastructure, perhaps building on recent work by \citet{sigal2021automatic} which shows how to express AD using effect handlers, or alternatively to interpret a model into a form suitable for use with an existing AD library such as \code{ad}~\cite{kmett2021ad}. In either case, we would need to investigate the set of valid operations for models in our language that are amenable to AD.

%In either case, we would need to investigate the set of valid operations or syntactic forms that enable an AD interpretation of models in our language.
% One way of incorporating AD into ProbFX is to interface with the Haskell compiler plugin by  which performs a source-to-source transformation of Haskell code.
% During this, we also hope to realise an approach to how handlers can be best performatively applied to PPLs.

\bibliographystyle{ACM-Reference-Format}
\bibliography{bibliography}
\newpage
\appendix
\section{Performance benchmarks}
\label{appendix:benchmarks}
\vspace{-0.65cm}
\begin{figure}[H]
  \hspace*{\fill}
  \begin{subfigure}{0.5\textwidth}

  \end{subfigure}
  \begin{subfigure}{0.4\textwidth}
    \includegraphics[width=1\columnwidth]{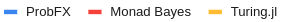}
  \end{subfigure}\\
  \begin{subfigure}{0.07\textwidth}
    \begin{adjustbox}{addcode={\begin{minipage}{2.7cm}}{\caption*{%
      \textsf{\scriptsize Lin Regr}
      }\end{minipage}},rotate=90,center}
  \end{adjustbox}
  \end{subfigure}
  \begin{subfigure}{0.3\textwidth}
    \caption*{\scriptsize \textsf{Simulation}}
    \centering
    \includegraphics[width=1\columnwidth, height=2.35cm]{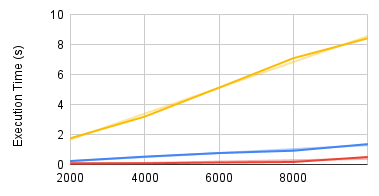}
    \vspace{-0.4cm}
  \end{subfigure}
  \hspace*{0.1cm}
  \begin{subfigure}{0.3\textwidth}
    \caption*{\scriptsize \textsf{Likelihood Weighting}}
    \center
    \includegraphics[width=1\columnwidth]{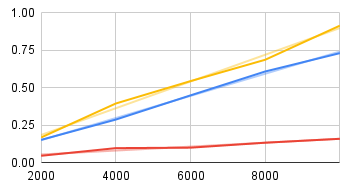}
    \vspace{-0.4cm}
  \end{subfigure}
  \begin{subfigure}{0.3\textwidth}
    \caption*{\scriptsize \textsf{Metropolis-Hastings}}
    \center
    \includegraphics[width=1\columnwidth]{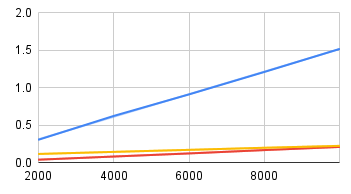}
    \vspace{-0.4cm}
  \end{subfigure} \\
  \begin{subfigure}{0.07\textwidth}
    \begin{adjustbox}{addcode={\begin{minipage}{2.7cm}}{\caption*{%
      \textsf{\scriptsize HMM}
      }\end{minipage}},rotate=90,center}
  \end{adjustbox}
  \end{subfigure}
  \begin{subfigure}{0.3\textwidth}
    \centering
    \includegraphics[width=1\columnwidth, height=2.7cm]{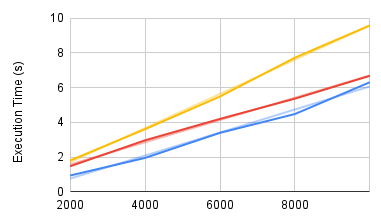}
    \vspace{-0.4cm}
  \end{subfigure}
  \hspace*{0.1cm}
  \begin{subfigure}{0.3\textwidth}
    \center
    \includegraphics[width=1\columnwidth]{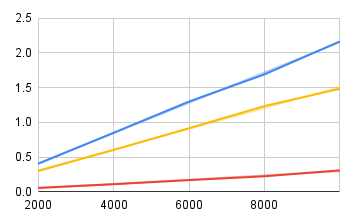}
    \vspace{-0.4cm}
  \end{subfigure}
  \begin{subfigure}{0.3\textwidth}
    \center
    \includegraphics[width=1\columnwidth]{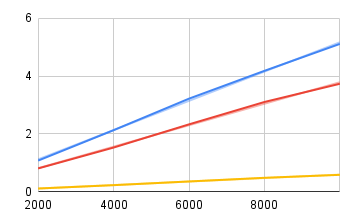}
    \vspace{-0.4cm}
  \end{subfigure} \\
  \begin{subfigure}{0.07\textwidth}
    \begin{adjustbox}{addcode={\begin{minipage}{2.7cm}}{\caption*{%
      \textsf{\scriptsize LDA}
      }\end{minipage}},rotate=90,center}
  \end{adjustbox}
  \end{subfigure}
  \begin{subfigure}{0.3\textwidth}
    \centering
    \includegraphics[width=1\columnwidth, height=3.1cm]{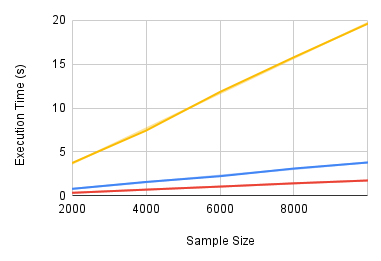}
    \vspace{-0.4cm}
  \end{subfigure}
  \hspace*{0.1cm}
  \begin{subfigure}{0.3\textwidth}
    \center
    \includegraphics[width=1\columnwidth]{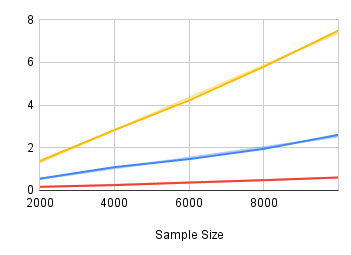}
    \vspace{-0.4cm}
  \end{subfigure}
  \begin{subfigure}{0.3\textwidth}
    \center
    \includegraphics[width=1\columnwidth]{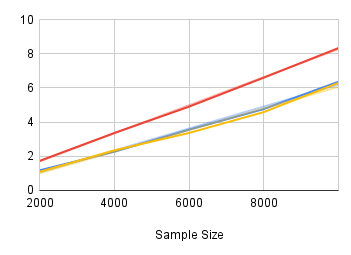}
    \vspace{-0.4cm}
  \end{subfigure}\\
  \vspace{-0.35cm}
  \caption{Execution times of algorithms (seen at top) applied to models (seen on left) with respect to the number of samples each algorithm generates (i.e. number of algorithm iterations).  }
  \label{fig:benchmark-sample-size}
\end{figure}

\vspace{-0.9cm}

\begin{figure}[H]
  \hspace*{\fill}
  \begin{subfigure}{0.5\textwidth}

  \end{subfigure}
  \begin{subfigure}{0.4\textwidth}
  \end{subfigure}\\
  \begin{subfigure}{0.07\textwidth}
    \begin{adjustbox}{addcode={\begin{minipage}{2.7cm}}{\caption*{%
      \textsf{\scriptsize Lin Regr}
      }\end{minipage}},rotate=90,center}
  \end{adjustbox}
  \end{subfigure}
  \begin{subfigure}{0.3\textwidth}
    \caption*{\scriptsize \textsf{Simulation}}
    \centering
    \includegraphics[width=1\columnwidth, height=2.35cm]{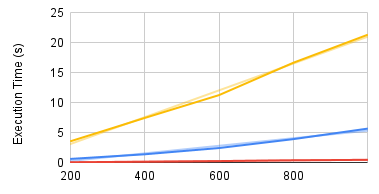}
    \vspace{-0.4cm}
  \end{subfigure}
  \hspace*{0.1cm}
  \begin{subfigure}{0.3\textwidth}
    \caption*{\scriptsize \textsf{Likelihood Weighting}}
    \center
    \includegraphics[width=1\columnwidth]{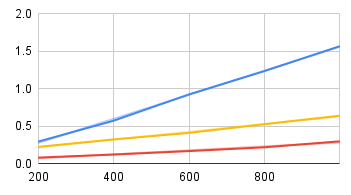}
    \vspace{-0.4cm}
  \end{subfigure}
  \begin{subfigure}{0.3\textwidth}
    \caption*{\scriptsize \textsf{Metropolis-Hastings}}
    \center
    \includegraphics[width=1\columnwidth]{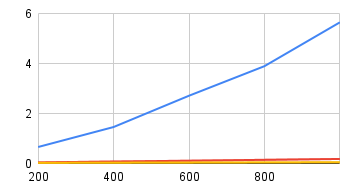}
    \vspace{-0.4cm}
  \end{subfigure} \\
  \begin{subfigure}{0.07\textwidth}
    \begin{adjustbox}{addcode={\begin{minipage}{2.7cm}}{\caption*{%
      \textsf{\scriptsize HMM}
      }\end{minipage}},rotate=90,center}
  \end{adjustbox}
  \end{subfigure}
  \begin{subfigure}{0.3\textwidth}
    \centering
    \includegraphics[width=1\columnwidth, height=2.5cm]{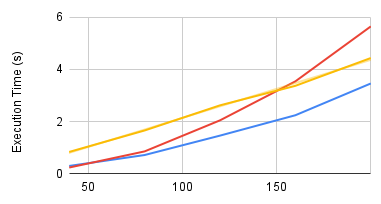}
    \vspace{-0.4cm}
  \end{subfigure}
  \hspace*{0.1cm}
  \begin{subfigure}{0.3\textwidth}
    \center
    \includegraphics[width=1\columnwidth]{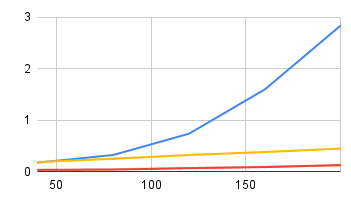}
    \vspace{-0.4cm}
  \end{subfigure}
  \begin{subfigure}{0.3\textwidth}
    \center
    \includegraphics[width=1\columnwidth]{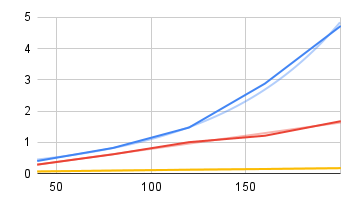}
    \vspace{-0.4cm}
  \end{subfigure} \\
  \begin{subfigure}{0.07\textwidth}
    \begin{adjustbox}{addcode={\begin{minipage}{2.7cm}}{\caption*{%
      \textsf{\scriptsize LDA}
      }\end{minipage}},rotate=90,center}
  \end{adjustbox}
  \end{subfigure}
  \begin{subfigure}{0.3\textwidth}
    % \vspace{-0.4cm}
    \hspace{-0.2cm}
    \centering
    \includegraphics[width=1.018\columnwidth, height=2.85cm]{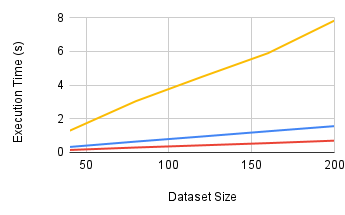}
    \vspace{-0.4cm}
  \end{subfigure}
  \hspace*{0.1cm}
  \begin{subfigure}{0.3\textwidth}
    \center
    \includegraphics[width=1\columnwidth]{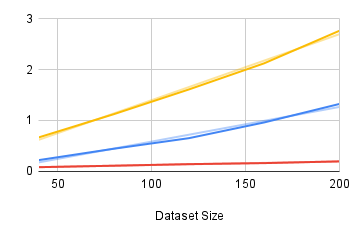}
    \vspace{-0.4cm}
  \end{subfigure}
  \begin{subfigure}{0.3\textwidth}
    \center
    \includegraphics[width=1\columnwidth]{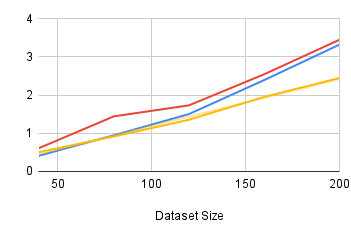}
    \vspace{-0.4cm}
  \end{subfigure}\\
  \vspace{-0.25cm}
  \caption{Execution times of algorithms (at top) applied to models (on left) with respect to dataset size provided to models. The data we vary over is: the number of data points in linear regression, the number of Markov nodes in the hidden Markov model, and the length of text document for latent dirichlet allocation.}
  \label{fig:benchmark-data-size}
\end{figure}

\end{document}